\documentclass[aps, prl, twocolumn, superscriptaddress, floatfix]{revtex4-2}
\usepackage{float}
\usepackage{hyperref}
\usepackage[utf8]{inputenc}
\usepackage{amsmath,amsfonts,amssymb}
\usepackage{bbold}
\usepackage{graphicx}
\usepackage{blkarray}
\usepackage{titlesec}
\usepackage{setspace}
\usepackage{color}

\newcommand{\bra}[1]{\left\langle #1\right|}
\newcommand{\ket}[1]{\left|#1\right\rangle}

\titleformat{\section}{\bfseries}{}{0em}{}[]
\titlespacing{\section}
  {0pt}
  {2.0ex plus 1ex minus 1ex}
  {3pt}
\titleformat*{\subsection}{\normalsize\itshape}
\titlespacing{\subsection}
  {0pt}
  {2.0ex plus 1ex minus 1ex}
  {3pt}

\begin{document}

\title{\vspace{-5ex} \large \textbf{Stimulated emission of signal photons from dark matter waves}}

\author{Ankur Agrawal}
\email{ankur.agrawal92@gmail.com, Present address: AWS Center for Quantum Networking, Boston, MA 02145, USA}
\affiliation{James Franck Institute, University of Chicago, Chicago, Illinois 60637, USA}
\affiliation{Department of Physics, University of Chicago, Chicago, Illinois 60637, USA}
\affiliation{Kavli Institute for Cosmological Physics, University of Chicago, Chicago, Illinois 60637, USA}

\author{Akash V. Dixit}
\email{Present address: National Institute of Standards and Technology, Boulder, CO 80305, USA}
\affiliation{James Franck Institute, University of Chicago, Chicago, Illinois 60637, USA}
\affiliation{Department of Physics, University of Chicago, Chicago, Illinois 60637, USA}
\affiliation{Kavli Institute for Cosmological Physics, University of Chicago, Chicago, Illinois 60637, USA}

\author{Tanay Roy}
\email{Present address: Superconducting Quantum Materials and Systems Center, Fermi National Accelerator Laboratory, Batavia, IL 60510, USA}
\affiliation{James Franck Institute, University of Chicago, Chicago, Illinois 60637, USA}
\affiliation{Department of Physics, University of Chicago, Chicago, Illinois 60637, USA}

\author{Srivatsan Chakram}
\email{Present address: Department of Physics and Astronomy, Rutgers, the State University of New Jersey, 136 Frelinghuysen Road, Piscataway, NJ 08854}
\affiliation{James Franck Institute, University of Chicago, Chicago, Illinois 60637, USA}
\affiliation{Department of Physics, University of Chicago, Chicago, Illinois 60637, USA}
\affiliation{Department of Physics and Astronomy, Rutgers University, Piscataway, New Jersey 08854, USA}

\author{Kevin He}
\affiliation{James Franck Institute, University of Chicago, Chicago, Illinois 60637, USA}
\affiliation{Department of Physics, University of Chicago, Chicago, Illinois 60637, USA}

\author{Ravi K. Naik}
\affiliation{Computational Research Division, Lawrence Berkeley National Laboratory, Berkeley, CA, USA}

\author{David I. Schuster}
\affiliation{James Franck Institute, University of Chicago, Chicago, Illinois 60637, USA}
\affiliation{Department of Physics, University of Chicago, Chicago, Illinois 60637, USA}
\affiliation{Pritzker School of Molecular Engineering, University of Chicago, Chicago, Illinois 60637, USA}
\affiliation{Department of Applied Physics, Stanford University, Stanford, California 94305, USA}

\author{Aaron Chou}
\affiliation{Fermi National Accelerator Laboratory, Batavia, Illinois 60510, USA}

\begin{abstract}
The manipulation of quantum states of light has resulted in significant advancements in both dark matter searches and gravitational wave detectors \cite{Dixit2021, Backes2021, Brubaker2017, Tse2019}. Current dark matter searches operating in the microwave frequency range use nearly quantum-limited amplifiers \cite{Braine2020, Brubaker2017, Kim2023}. Future high frequency searches will use photon counting techniques~\cite{Dixit2021} to evade the standard quantum limit. We present a signal enhancement technique that utilizes a superconducting qubit to prepare a superconducting microwave cavity in a non-classical Fock state and stimulate the emission of a photon from a dark matter wave. By initializing the cavity in an $\ket{n=4}$ Fock state, we demonstrate a quantum enhancement technique that increases the signal photon rate and hence also the dark matter scan rate each by a factor of 2.78. Using this technique, we conduct a dark photon search in a band around $\mathrm{5.965\, GHz \, (24.67\, \mu eV)}$, where the kinetic mixing angle $\epsilon \geq 4.35 \times 10^{-13}$ is excluded at the $90\%$~confidence level.
\end{abstract}

\maketitle
\section*{Introduction}
The existence of dark matter (DM) is one of the greatest mysteries in physics, which has puzzled scientists for nearly a century. Despite the lack of direct detection, there is compelling evidence for its existence, including its estimated contribution of 27\% to the universe's energy density and its gravitational effects on galaxy dynamics and structure formation \cite{Chou2022, Tanabashi2018, Rubin1980}. Axions and dark photons have emerged as leading candidates for dark matter due to their cosmological origins and low-energy properties, which allow them to exist as coherent waves with macroscopic occupation numbers \cite{Preskill1983, Abbott1983, Dine1983, Arias2012, Graham2016}. Dark matter haloscope experiments in the microwave frequency range use a cavity to resonantly enhance the oscillating electric field generated by the DM field at a frequency corresponding to the mass of a hypothetical particle ($\nu = mc^{2}/h$)\cite{Sikivie1983, Graham2016}. Since the mass of DM is unknown a priori, experimental searches are typically conducted as radio scans in which the resonant cavity is tuned one step  $d\nu$ at a time to test different frequency hypotheses. A key figure of merit is therefore the frequency scan speed, which in a photon counting experiment scales as $d\nu / dt \propto d\nu\,  R_{s}^2/R_{b}$ where $R_{s}$ and $R_{b}$ are the signal and background count rates respectively.

Quantum techniques have proven useful in accelerating the scan rate of axion and wavelike dark matter searches. Superconducting parametric amplifiers, which are quantum-limited, have reached the standard quantum limit (SQL) and add only 1/2 photon of noise per mode, as required by the Heisenberg uncertainty principle for phase-preserving measurements \cite{Caves1982, Yamamoto2008, Eichler2014, Tanay2015, Martina2019}. Alternatively, qubit-based single photon detection \cite{Dixit2021} does not consider the photon phase and can in principle measure without any added detection noise by achieving extremely low, sub-SQL background rates.

While both of these methods employ quantum technologies, their operation is in some sense recognizable as an ideal classical amplifier and microwave photo-multiplier. Parametric amplifiers and cavity-qubit systems can also synthesize inherently quantum mechanical states of light such as squeezed states \cite{Caves1981, Lawrie2019, Eickbusch2022} in the former and Fock states \cite{Hofheinz2008, Wolf_2019}  or Schrodinger cat states \cite{Schrodinger1935, Heeres2017, Ofek2016, Hu2019, Campagne_Ibarcq2020} for the latter.  Recently, squeezed state injection paired with phase-sensitive amplification was used to improve the scan rate of the HAYSTAC experiment \cite{Backes2021}. In this work, we develop a new method in which we prepare an $n$-photon Fock state which enhances the signal rate by $\eta\,(n+1)$ times where $\eta$ is the efficiency of detecting the state. By creating a Fock state with $\ket{n=4}$ photons in the cavity, we observe a $2.78$-fold enhancement in the signal rate. We show that this technique is compatible with the previously-demonstrated noise reduction from photon counting \cite{Dixit2021}.

The power delivered to the cavity by a current density generated by dark matter $\textbf{J}_{DM}$ is given by
\begin{equation}
    P_{s} =  \int dV \, \textbf{J}_{DM}(x) \cdot \textbf{E}(x)
    \label{eq:pow_axion}
\end{equation}
and is proportional to the magnitude of oscillating electric field $\textbf{E}(x)$ in the cavity. In the conventional scenario, the cavity is cooled to the vacuum state, and the signal electric field builds up monotonically over the coherence time of the cavity or of the dark matter wave in a process akin to spontaneous emission.  Alternatively, we may initialize the cavity with a non-zero $\textbf{E}(x)$ field from a coherent state sine wave or from a Fock state to induce stimulated emission.  The Fock state has some advantages.  First, unlike the homodyne or heterodyne detection using a coherent state pump, the Fock state is free from any shot noise, making it possible to measure small signal amplitudes far below the SQL. Second, a Fock state is symmetric in phase, making it equally sensitive to any instantaneous phase of the incoming DM wave, which is unknown a priori.  

We can model the action of the DM wave on a Fock state as a classical drive amplitude $\xi$ which shifts this phase-symmetric state away from the origin in the Wigner phase space by $\alpha$ (see Supp Fig. \ref{fig:stim_em_concept}). The resultant state comprises both in-phase components which extracted excess power from the DM wave and also out-of-phase components which delivered their power to the DM wave. The stimulated emission process for DM converting into photons is enhanced by a factor of $(n+1)$ while the stimulated absorption process is enhanced by a factor of $n$. Mathematically, the stimulated emission into the cavity state from a dark matter wave can be described as shown below,
\begin{equation}
    \begin{split}
    |\bra{n+1}  \mathcal{\hat{D}(\alpha)} \ket{n}|^{2} & = |\bra{n+1} e^{(\alpha a ^ {\dagger} - \alpha^{*} a)} \ket{n}|^{2} \\
    & \sim |\bra{n+1} \alpha a^{\dagger} \ket{n}|^{2}  = (n+1) \alpha^{2}
    \end{split}
    \label{eqn:stim_em}
\end{equation}
where $\mathcal{\hat{D}(\alpha)}$ is the displacement operator. From Eqn. \eqref{eqn:stim_em}, we can infer that the displacement ($\alpha \ll 1$) induced by the DM wave on a cavity prepared in $\ket{n}$ Fock state results in population of $\ket{n+1}$ state with probability proportional to $(n+1)$. Using number-resolving measurements, the signal can thus be observed with $(n+1)\times$ higher probability if the cavity is prepared in a larger $\ket{n}$ Fock state.

We note that just as in other cases of quantum-enhanced metrology, the $(n+1)$ enhancement factor in the signal transition probability can be exactly canceled by the $1/(n+1)$ reduction in the coherence time of the probe. As a result, there would be no net improvement in the actual signal rate $R_{s}$. However, this consideration does not apply when the limiting coherence time is that of the dark matter wave rather than of the Fock photon state in the cavity. In such cases, the signal rate is not degraded by the reduction of the probe coherence time and retains the factor $(n+1)$.  To our knowledge, this is one of the few cases where quantum metrology can provide a realizable improvement in a real-world application. Also, since the readout rate scales as the inverse of the probe coherence time, the background count rate may also scale linearly with $(n+1)$, for example for backgrounds associated with readout errors.  For experiments with fixed tuning step size $d\nu$ given for example by the dark matter linewidth, the improvement in scan speed $d\nu / dt \propto d\nu\, R_{s}^2/R_{b}$ is therefore a single factor of $\eta \, (n+1)$.  

\section {Fock state preparation and photon number resolving detector}
\begin{figure}[ht]
    \centerline{
    \includegraphics[width=\columnwidth]{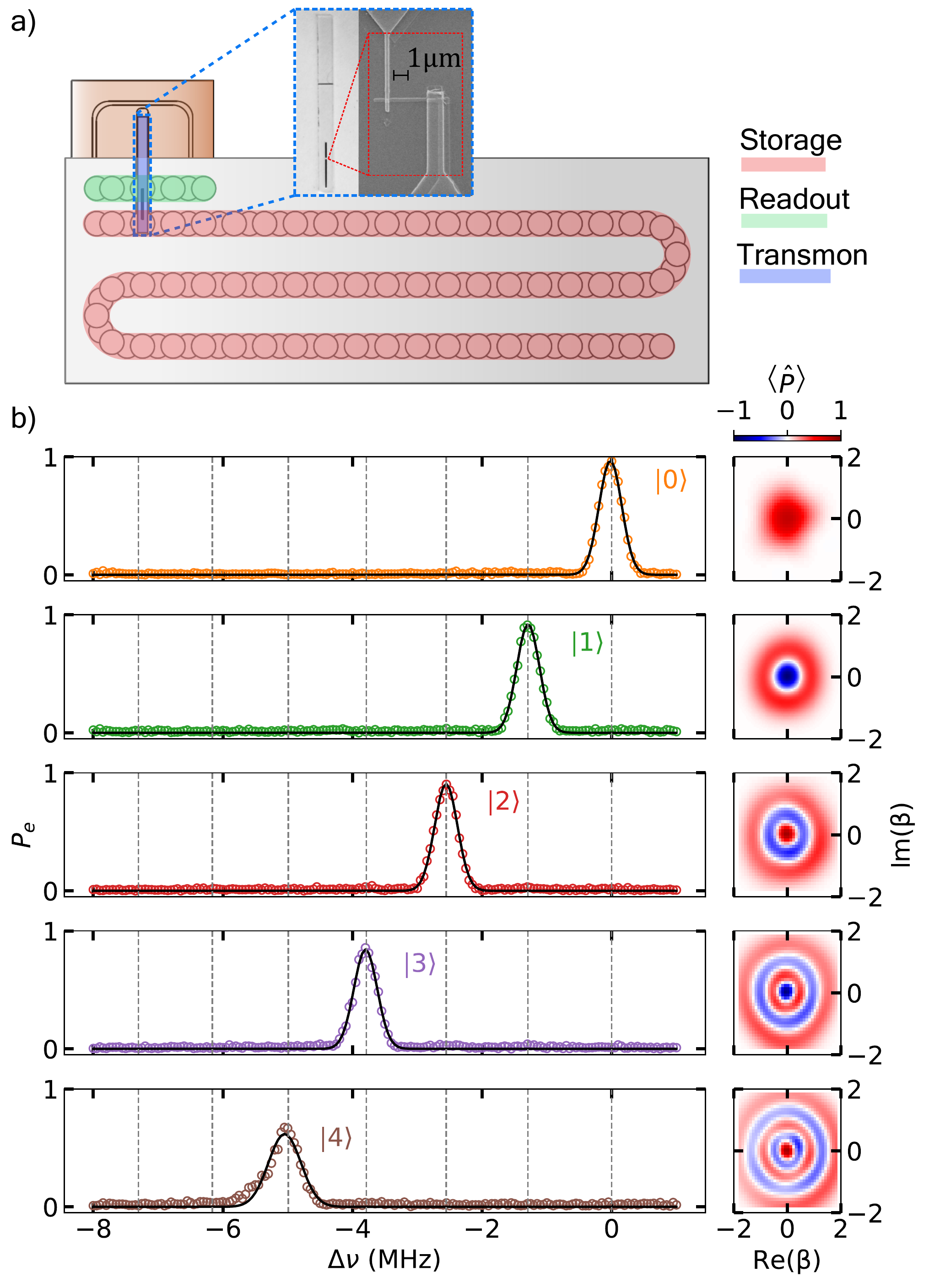}
    }
    \caption{\textbf{Fock state preparation in a cavity dispersively coupled to a transmon qubit.} (a) A schematic of the multimode flute cavity showing the location of the storage cavity (red), readout cavity (green), and transmon chip with a SEM image of the Josephson junction (blue) \cite{Chakram2021}. (b) Creation of Fock states in a particular mode of the cavity using GRAPE method. Characterization of the cavity state using qubit spectroscopy (left) and Wigner tomography (right). Qubit spectroscopy is performed immediately after the optimal control (OCT) pulses, a single peak in each qubit excitation probability ($P_{e}$) distribution confirms the creation of the correct $\ket{n}$ Fock state. Resultant probability distribution is fitted to a Gaussian to estimate the state preparation fidelity. Grey dashed lines correspond to the expected shift in frequency, accounted for quadratic dispersive shift. (Right) Wigner tomography \cite{Cahill1969} is performed by coherently displacing the resultant cavity state in 2D phase space to map the average parity and thus, reconstruct the cavity state density matrix.}
    \label{fig:device_spec_wt}
\end{figure}

We couple the cavity to a non-linear element, in this case a superconducting transmon qubit to prepare and measure the Fock states, which are otherwise impossible to create in a linear system such as a cavity. The device used in this work is composed of three components - a high quality factor ($Q_s = 4.06 \times 10^7$) 3D multimode cavity \cite{Chakram2021} to accumulate and store the signal induced by the dark matter (storage, $\omega_s/ 2 \pi = \mathrm{5.965\, GHz}$), a superconducting transmon qubit ($\omega_q /2 \pi = \mathrm{4.95\, GHz}$), and a 3D cavity strongly coupled to a transmission line ($Q_r = 9 \times 10^3$) used to quickly read out the state of qubit (readout, $\omega_r / 2 \pi = \mathrm{7.789\, GHz}$) (Fig. \ref{fig:device_spec_wt} (a)). We mount the device to the base stage of a dilution refrigerator operating at $\mathrm{10\, mK}$.

The interaction between a superconducting transmon qubit \cite{Koch2007, Ambegaokar1963} and the field in a microwave cavity is described by the Jaynes-Cummings Hamiltonian \cite{Jaynes_1963} in the dispersive limit (qubit-cavity coupling $\ll$ qubit-cavity detuning) as, 
\begin{equation}
\mathcal{H}/\hbar = \omega_s a^{\dagger} a + \frac{1}{2}(\omega_q + \chi a^{\dagger} a) \sigma_z
  \label{eqn:hamiltonian}
\end{equation}
where $a$ and $a^{\dagger}$ are the annihilation and creation operators of the cavity mode and $\sigma_z$ is the Pauli $Z$ operator of the transmon. Eqn. \ref{eqn:hamiltonian} elucidates a key feature of this interaction - a photon number dependent shift ($\chi$) of the qubit transition frequency (see Fig. \ref{fig:device_spec_wt}(b))\cite{Schuster2007}. Another important feature of this Hamiltonian is the quantum non-demolition (QND) nature of the interaction between the qubit and cavity which preserves the cavity state upon the measurement of the qubit state and vice-versa \cite{Brune1990, Gleyzes2007, Schuster2007}. By driving the qubit at the Stark shifted frequency ($\omega_{q} + n \,\chi $), one would selectively excite the qubit if and only if there are exactly $n$-photons in the cavity.

Recent works have shown that a single transmon has the capability to prepare any quantum state in a cavity and perform universal control on it  \cite{Heeres2015, Heeres2017, Wang2008, Nelson2017, Chakram2022, Eickbusch2022}. In this study, we used a GRadient Ascent Pulse Engineering (GRAPE) based method to generate optimal control pulses (OCT) \cite{Khaneja2005, Heeres2017} that consider the full model of the time-dependent Hamiltonian and allow us to prepare non-classical states in a cavity. As shown in Fig. \ref{fig:device_spec_wt} (b), our approach successfully prepared cavity Fock states with pulse duration as short as $\mathcal{O}(1/ \chi)$, which did not increase for higher Fock states. 

\section{Stimulated emission protocol}
\begin{figure}[ht]
    \centerline{
    \includegraphics[width=\columnwidth]{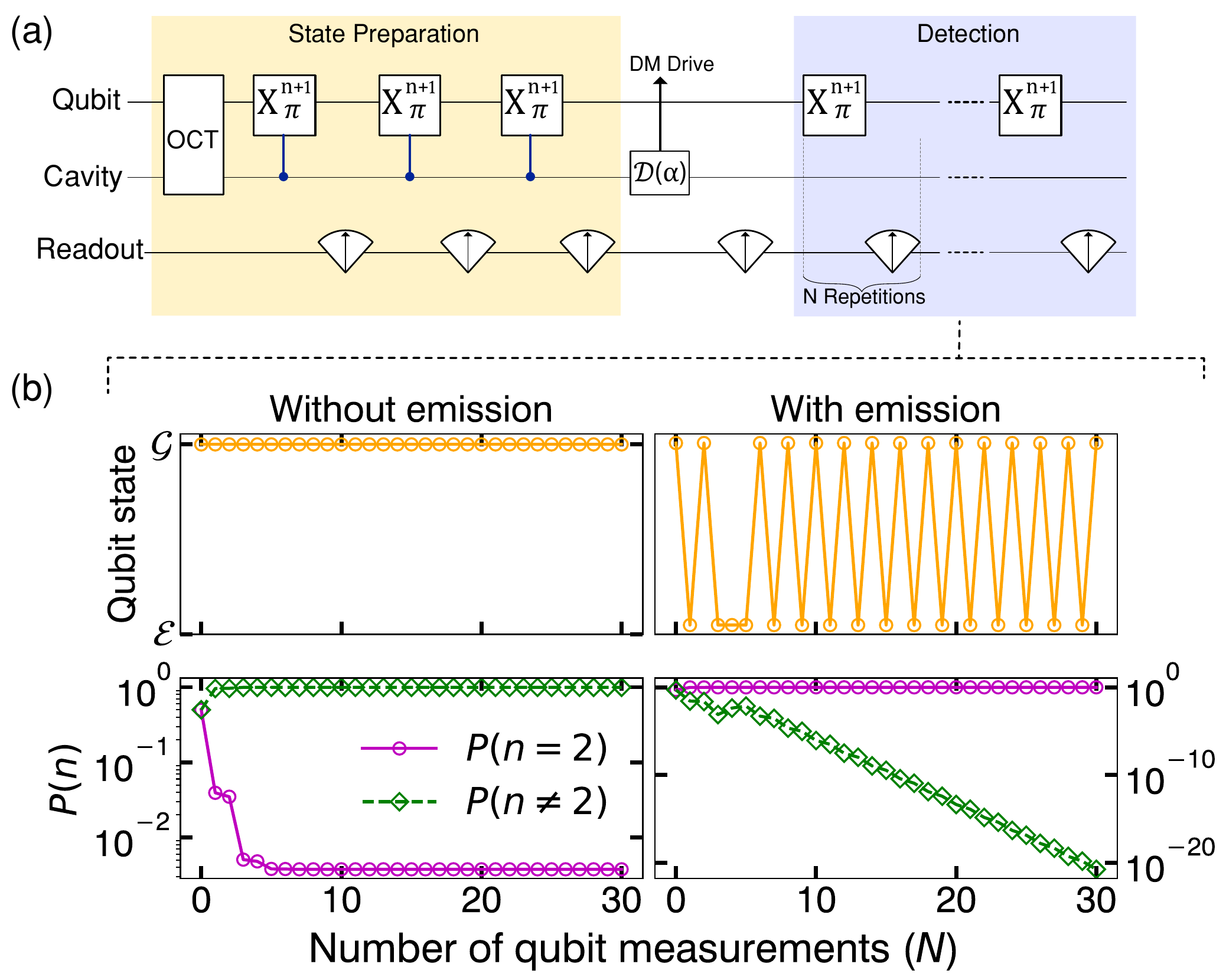}
    }
    \caption{\textbf{Stimulated emission protocol with number resolved $\pi$-pulse and hidden Markov model analysis.} (a) Pulse sequence for stimulated emission includes cavity initialization in a Fock state, followed by three conditional checks to ensure cavity did not accidentally start in $\ket{n+1}$ state. The next part involves a cavity displacement drive to mimic a push from the DM and repeated conditional qubit measurements to detect the cavity in $\ket{n+1}$ state, where the first measurement collapses the cavity state to either $\ket{n+1}$ or not. If in $\ket{n+1}$, the subsequent measurements are QND and via the quantum Zeno effect, each measurement resets the clock on the decay of the $\ket{n+1}$ state. (b) Examples showing two measured qubit readout sequences for a cavity initialized in $\ket{n}=1$ Fock state followed by a small displacement drive $\alpha$. The left panel corresponds to no change in the cavity state after the DM drive as inferred by the absence of successful flips of the qubit state which results in a very small probability $P(n=2)$ that the cavity was in the $\ket{n}=2$ Fock state. The right panel corresponds to an emission event where the cavity state changed from $\ket{1}\xrightarrow{}\ket{2}$, resulting in multiple successful flips of the qubit state.  The HMM analysis of this sequence of flips then indicates a very high likelihood ratio to be in $\ket{n}=2$ Fock state. In case of successful detection, we observe an exponential suppression of the detector error based false positive probability with only linear increase in the number of repeated measurements.}
    \label{fig:pulse_meas}
\end{figure}

The stimulated emission protocol is divided into two parts: the first part involves the  preparation of cavity in a desired Fock state, $\ket{n}$ and the second part involves the detection of the cavity in the $\ket{n+1}$ Fock state as depicted in Fig. \ref{fig:pulse_meas}(a). In order to actively suppress any false positive events such as the cavity accidentally starting in $\ket{n+1}$ state, we conditionally excite the qubit with a resolved $\pi$-pulse at the $(n+1)$-shifted peak three times. If and only if the qubit fails to excite in all three attempts do we proceed ahead with rest of the protocol. By doing this, we can suppress the false positive rate $\leq 3\%$. 

At the end of this sequence, we measure the efficiency of the state preparation for each $n$ by measuring the qubit excitation probability $P_{n}$ with a number resolved $\pi$-pulse centered at the $\ket{n}$ peak. The measured fidelities are $P_{0} = 95.2\pm0.3\%, P_{1} = 91.2\pm0.4\%, P_{2} = 87.3\pm0.5\%, P_{3} = 81.6\pm0.6\%, P_{4} = 63.6\pm0.7\%$. 

After the state preparation, we apply a coherent drive to the cavity mimicking a push from the DM wave to characterize the detector. A series of repeated QND measurements are recorded by performing conditional $\pi$-pulses centered at the $(n+1)$-shifted peak. The time between two successive QND measurements is $\mathrm{5\, \mu s}$, which is relatively short compared to the lifetime of Fock states given by $T_{1}^{s} = \mathrm{1320\, \mu s}/n$ (see Table \ref{table:qndness}). This projective measurement resets the clock on the decay of the $\ket{n+1}$ state \cite{Itano1990}. We then apply a hidden Markov model (HMM) analysis to reconstruct the cavity state and compute the probability that the cavity state had changed from $\ket{n} \rightarrow \ket{n+1}$  and assign a likelihood ratio $\lambda$ associated with such events (see Supplementary section for implementation of HMM analysis).

In principle, it is possible to prepare Fock states $\ket{n>4}$ in the device. However, due to the presence of multiple cavity modes, simulating the complete Hamiltonian to generate the OCT pulses becomes computationally challenging and in practice, higher $(n+1)$ Fock states are prepared with lower fidelity. Furthermore, to prevent excessive signal photon loss, the Fock state decay rate which is also enhanced by a factor of $(n+1)$ must remain small compared to the sum of Stark shift and readout rate, which determines the maximum rate of number resolved measurements. For this study, we chose $\ket{n=4}$, such that the decay probability stays below $1\%$ between successive measurements.

\section{Signature of Fock enhancement}
To assess the performance of the detector after preparing the cavity in a particular Fock state $\ket{n}$, we carry out a series of experiments. We apply a small variable displacement ($\alpha \ll 1$) to the cavity and measure the relationship between the number of injected ($n_{\rm inj} = |\alpha|^{2}$) and detected photons. We perform 30 repetitions of the qubit measurement and apply a likelihood threshold of $\lambda_{\mathrm{thresh}}=10^{3}$ to distinguish positive and negative events. This threshold is determined based on the background cavity occupation $n_{b}^{c}=6\times10^{-3}$, which is assumed to be caused by photon shot-noise from a hot cavity, as measured using the photon counting method described in \cite{Dixit2021} (refer to Fig. \ref{fig:cavity_bkgd_pc}). Errors below this value are considered to be sub-dominant. For a cavity initialized in $\ket{n}$, the probability of finding the cavity in the Fock state $\ket{l}$ for a complex displacement $\alpha$ is described by the analytical expression \cite{de_Oliveira1990}: $P_{nl} (|\alpha|^{2}) = \big| \bra{l}\hat{\mathcal{D}}(\alpha)\ket{n}\big|^{2} = (n!/l!) \alpha^{2(l-n)} e^{-|\alpha|^{2}} \times \mathcal{L}{n}^{l-n}(|\alpha|^{2})$, where $\mathcal{L}{n}^{l-n}$ is an associated Laguerre polynomial. The data obtained from the characterization of the detector is fitted to an expression, represented by Eqn. \eqref{eq:stime_em_fit}. 
\begin{equation}
    n_{\rm meas} = \eta\, P_{nl} (|\alpha|^{2} = \bar{n}) + \delta
    \label{eq:stime_em_fit}
\end{equation}
This equation takes into account the detection efficiency, $\eta$, and false positive probability, $\delta$. In cases where the cavity displacement $\alpha \ll 1$, the relationship between the injected photons ($n_{\rm inj} = |\alpha|^{2}$) and measured photons can be approximated such that $P_{nl} (|\alpha^{2}|) \approx (n+1)|\alpha^{2}|$, as shown in Eqn. \eqref{eqn:stim_em}

\begin{figure}[ht]
    \centerline{
    \includegraphics[width=\columnwidth]{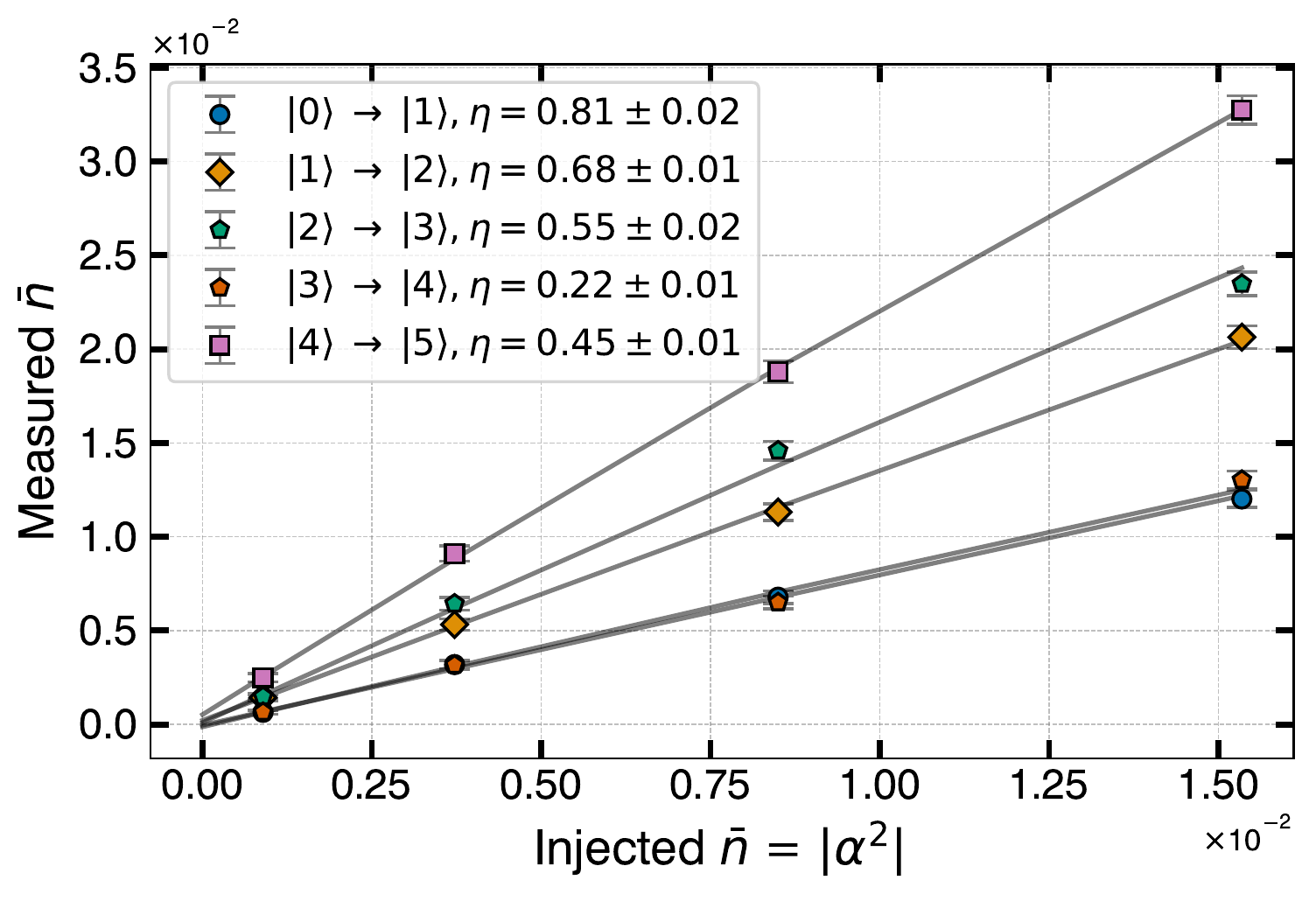}
    }
    \caption{\textbf{Stimulated emission enhancement}. Mean number of measured photons (positive events) as a function of the mean number of injected photons in the cavity. After initializing the cavity in a Fock state $\ket{n}$ and applying a variable cavity displacement (mock DM drive), 30 repeated qubit measurements of the cavity photon state are performed and a threshold $\lambda_{\mathrm{thresh}}$ is applied to determine the cavity population at $\ket{n+1}$. Background events with $\alpha=0$ are subtracted to compare between different Fock states which may have systematic errors in the state preparation step. $\lambda_{\mathrm{thresh}} = 10^{3}$ is chosen based on the observed background cavity occupation of $n_{b}^{c}=6\times10^{-3}$ such that the detector based errors are still sub-dominant. The background cavity occupation is measured using the photon counting technique described in Ref. \cite{Dixit2021} with repeated qubit parity measurements. Detector efficiency ($\eta$) for each Fock state is determined from the fit as reported in the legend. The monotonic decrease in the efficiency is attributed to  - higher decay probability to reach the same false positive probability and demolition probability (see Supp Fig. \ref{fig:qndness_storage}). Anomalous behavior in $\ket{3}$ is attributed to the state decaying to nearby modes in the band structure formed by multiple modes of the cavity, qubit and readout, which are close in the energy level $\ket{q, s, r}$.}
    \label{fig:det_char_stimem}

\end{figure}

Fig. \ref{fig:det_char_stimem} displays the unique characteristic of stimulated emission enhancement, where a higher number of detected photons is observed for the cavity that was initialized in a higher Fock state. This result aligns with expectations and highlights the effectiveness of stimulated emission as a method for amplifying weak signals. The figure also showcases a clear, monotonic increase in the slope as the Fock states increase, further emphasizing the success of this technique. The resultant enhancement between $\ket{n=4}$ and $\ket{n=0}$ is 2.78 ($0.45\times 5/0.81 \times 1$). The reduced efficiency $\eta=0.45$ to see the full $(n+1)$ enhancement can be explained by the enhanced decay rate of higher Fock states making it more difficult to achieve the fixed likelihood ratio threshold used for the comparison.  For example, consider the $\ket{n=5}$ Fock state. During the course of 30 repeated measurements, it is 1.6 times more likely to decay than the $\ket{n=1}$ Fock state. In addition, the higher demolition probability 0.074 for $\ket{n=4}$ makes it more difficult to achieve high likelihood ratio since this photon state only persists for the first 14 quasi-QND measurements. The false positive probabilities $\delta$ are smaller than $10^{-4}$ for all Fock states, comparable to the measured residual photon occupation in the cavity. Further advancements could be achieved by utilizing a system with a higher Q value and reduced demolition probability. By combining weakly coupled qubits with the Echoed Conditional Displacement (ECD) technique \cite{Eickbusch2022}, Fock states can be prepared with high fidelity while also keeping the demolition probability low. Additionally, a detector with reduced errors from factors such as thermal population and readout fidelity would also improve the results.

We observe anomalous behavior for the $\ket{n}=3$ data which shows no signal enhancement. There is no direct way to investigate the cause but we suspect leakage to a nearby mode as the 3D cavity contains multiple modes which are closely spaced \cite{Chakram2021}. We tried this experiment on a different cavity mode and observed similar behavior but for $\ket{n}=1$. We have identified a couple of transitions with different modes which are closer in energy level with $\ket{g, 3}$ and which could be facilitated by the always-on interaction of the transmon with all the modes. This frequency collision issue can be easily resolved in future designs such that the cavity modes are spaced further apart and the transmon has negligible overlap with the spectator modes.

\section{Dark photon search}
In order to conduct a dark photon search, we collect independent datasets for a cavity prepared in different Fock states and count the number of positive events in the absence of the mock DM drive. Additionally, we vary the dwell time ($\tau$) between the state preparation and the beginning of the measurement sequence in order to allow the coherent buildup of cavity field due to the dark matter. Once the measurement sequence begins, the quantum Zeno effect prevents further build-up of the signal field. Ideally, one would like to choose the dwell time as close to the lifetime of Fock state as possible to maximize the accumulation of signal and thus, the scan rate. In this work, the dwell time was varied to compare on equal footing the dependence on $n$ and not optimized for DM sensitivity. For this study, we chose a maximum dwell time of $\mathrm{20\, \mu s}$ to collect reasonable statistics with $\lambda_{\mathrm{thresh}} = 10^{3}$ for all Fock states. Longer integration times comparable to or larger than the dark matter coherence time ($\mathrm{75\, \mu s}$) which are needed to realize the full benefit of Fock enhancement will be implemented in future dedicated experiments.
\begin{figure}[h]
    \centerline{
    \includegraphics[width=\columnwidth]{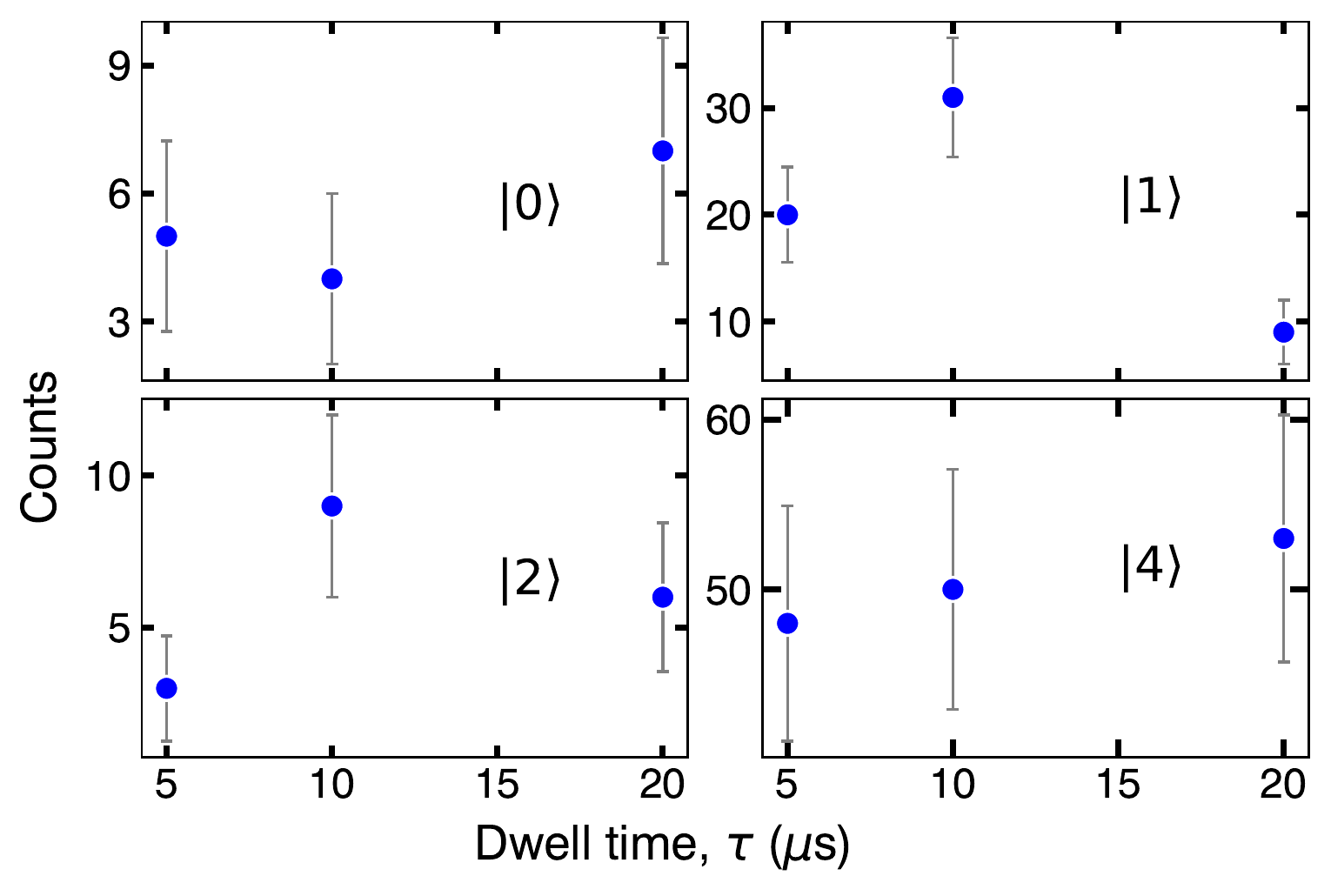}
    }
    \caption{\textbf{Measured background counts for different Fock states in the cavity as a function of dwell time.} There is no clear trend in the number of observed background counts indicating systematic effects which could be due to the state preparation steps. The error bars are plotted as $\sqrt{\mathrm{Counts}}$ and the $N_{\mathrm{trials}} \sim 20, 000$ for each point.}
    \label{fig:cavity_bkgd_dwell}
\end{figure}
The number of measured counts shown in Fig. \ref{fig:cavity_bkgd_dwell} is fit to a functional form given by Eqn. \eqref{eq:hp_fit_eqn} (see Supplementary material), which has contributions coming from a coherent source (hence the ($n+1$) Fock enhancement factor), an incoherent source,  and a state preparation dependent error
\begin{multline}
    N_{\rm meas} = a_{0}\, (n+1)\, \tau \, (N_{\rm trials}\, \tau) + b\, (N_{\rm trials}\, \tau) + \\ c_{n}\, N_{\rm trials}
    \label{eq:hp_fit_eqn}
\end{multline}
where $a_{0}$ and $b$ and $c_{n}$'s are the fit parameters we extract from fitting the measured counts. The first term has two factors of $\tau$: one from the coherent buildup of signal energy in the storage cavity for $\tau < T_{1}^{s}$ which is included in the average signal rate $dN/dt$, and a second factor of $\tau$ for the total integration time $t_{\rm tot}= N_{\rm trials} \tau$.

\begin{figure}[htbp]
    \centerline{
    \includegraphics[width=\columnwidth]{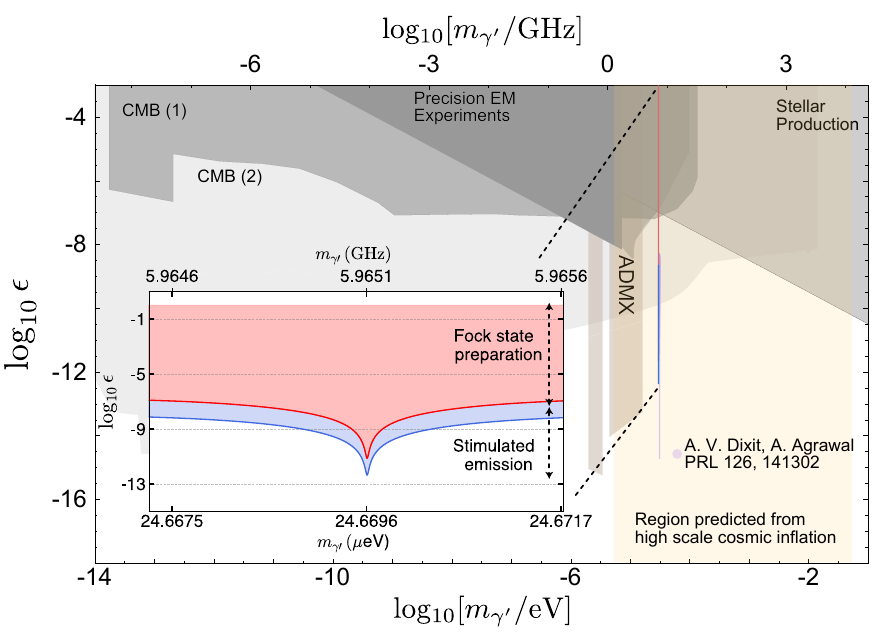}
    }
    \caption{\textbf{Exclusion of dark photon parameter space with stimulated emission}. Shaded regions in the dark photon parameter space \cite{Arias2012, McDermott2020} of coupling ($\epsilon$) and mass ($m_{\gamma}$) are excluded. In the orange band, dark photon is naturally produced in models of high scale cosmic inflation \cite{Graham2016}. The exclusion set with stimulated emission based dark photon search is shown in the blue and red curves. On resonance with the storage cavity ($m_{\gamma '} c^2 = \hbar \omega_s$), the dark photon kinetic mixing angle is constrained to $\epsilon \leq 4.24 \times 10^{-13}$ with 90\% confidence. (Inset) The horizontal extent of the excluded region is set by the bandwidth of the number resolved qubit $\pi$-pulse which is insensitive to any drive outside the band. The vertical limit is set by the maximum $\epsilon$ which would result in dark photon rate greater than the value which would degrade the fidelity of Fock state preparation significantly. The region between the blue and red curve represents the exclusion with the stimulated emission experiment, whereas the excluded region above the red curve is mainly due to the failure of Fock state preparation which is easily detectable in the experiment (See Supp Fig. \ref{fig:max_alpha}).}
    \label{fig:limit_plot}
\end{figure}
By performing an ordinary least square (OLS) fit to the measured counts, we extract the fit parameters with their uncertainties tabulated in Table \ref{table:fit_params_bkgd}. The values of fit parameters obtained from performing a Maximum Likelihood Estimate (MLE) were comparable.
\begin{table}[h]
\begin{center}
\begin{tabular}{ c c c }
\hline
 Fitted Parameter & $\Theta$ & $\sigma_{\Theta}$ \\
 \hline
 $a_{0}\, (\mathrm{s^{-2}})$ & $1.9 \times 10^{3}$ & $7.662\times 10^{5}$ \\
 $b\, (\mathrm{s^{-1}})$  & $-7.26$ & $4.2\times 10^{1}$ \\
 $c_{0}$ & $3.402\times 10^{-4}$ & $4.292\times 10^{-4}$ \\
 $c_{1}$ & $1.419\times10^{-3}$ & $4.0 \times 10^{-4}$ \\
 $c_{2}$ & $5.860\times 10^{-4}$ & $4.222\times 10^{-4}$ \\
 $c_{4}$ & $7.330\times 10^{-3}$ & $6.732\times 10^{-4}$ \\
 \hline
\end{tabular}
\caption[Fitted parameters.]{\textbf{Fitted parameters.} Best fit and statistical uncertainties corresponding to each source of background counts. The Fock state with higher background counts have larger $c$ value indicating that the source of counts is related to the state preparation step and thus, it is valid to perform a background subtraction to demonstrate the enhancement technique. For $\tau = \mathrm{20\, \mu s}$ and $N_{\mathrm{trials}} \sim 20, 000$, all three terms in Eqn. \eqref{eq:hp_fit_eqn} contribute roughly equally to the counts observed in Fig. \ref{fig:cavity_bkgd_dwell}.}
\label{table:fit_params_bkgd}
\end{center}
\end{table}

The large statistical uncertainties on $a_{0}$ are due to the fact that the other two terms dominate the measured counts and fluctuations causing $a_{0}$ to swing up and down by a large amount. With the extracted value of $a_{0}$, we can compute the kinetic mixing angle of dark photon given by $ \epsilon = \sqrt{\frac{a_{0}}{\rho_{\mathrm{DM}} m_{\mathrm{DM}} G V}}$ (see Supplemental material for derivation). With the measured uncertainties on all the parameters and using error propagation, we compute the $90\%$ confidence limit on $\epsilon$ to be $\epsilon_{fit} + 1.28\, \sigma_{\epsilon}$. Thus, a dark photon candidate on resonance with the storage cavity ($m_{\gamma '} c^2 = \hbar \omega_s$), with mixing angle $\epsilon \geq 4.35 \times 10^{-13}$ is excluded at the 90\% confidence level. Fig. \ref{fig:limit_plot} shows the regions of dark photon parameter space excluded by the stimulated emission based search, assuming the dark photon comprises all the dark matter density ($\rho_{\mathrm{DM}} = \mathrm{0.4\, GeV / cm^{3}}$). The detector is maximally sensitive to dark matter candidates with masses within a narrow window around the resonance frequency of the cavity. This window is set by the lineshape of the dark matter \cite{Foster2018} ($Q_{\mathrm{DM}} \sim 10^6$). For DM wave at $\sim \mathrm{6\, GHz}$, the FWHM linewidth is $\mathrm{6\, kHz}$, so the -$\mathrm{3\, dB}$ point is $\mathrm{3\, kHz}$ away from the cavity resonance and the efficiency of a finite bandwidth number resolved qubit $\pi$-pulse is non-zero. While the stimulated emission protocol is valid for small induced signal, the cavity state preparation using OCT pulses excludes DM wave corresponding to $|\alpha| > 0.05$ (see Fig. \ref{fig:max_alpha}) as it would result in larger preparation error than experimentally observed.  
\section{Conclusions and outlook}
The preparation of a cavity in quantum states of light, such as a Fock state, has the potential to significantly enhance the DM signal rate and hence also the DM frequency scan rate compared to previous methods. In the current work, we have demonstrated a signal enhancement of $2.78 \times$ by initializing the cavity in $\ket{n=4}$ versus $\ket{n=0}$ Fock state. The corresponding improvement in attainable scan speed is reduced because of the finite amounts of time required to prepare the initial Fock state and to read out the final state after integrating the signal, though in an optimized experiment the duty cycle should be comparable to that of other photon counting experiments.  Despite these imperfections, even in this initial demonstration, the dark photon search sets an unprecedented sensitivity in an unexplored parameter space with $90\%$ confidence level. This method holds great promise for continued improvement as advancements in cavity coherence times \cite{Milul2023} and state preparation methods \cite{Eickbusch2022} progress, and we expect the same improvement factor or better in future dedicated dark matter search experiments. 

While this study focuses on the detection of dark matter, the quantum-enhanced technique presented here can be applied more widely to sense ultra-weak forces in various settings, in cases where the signal accumulation is limited by the coherence time of the signal rather than by that of the probe. The Fock state stimulated emission can increase the rate of processes that involve the delivery of small amounts of energy, and the resulting signal quanta can be detected through number-resolved counting techniques that surpass the SQL. 

\section{Acknowledgements}
We would like to acknowledge and thank Konrad Lehnert for proposing this idea. We thank Ming Yuan, A. Oriani, and A. Eickbusch for discussions, and the Quantum Machines customer support team for help with implementing the control software. We gratefully acknowledge the support provided by the Heising-Simons Foundation. This work made use of the Pritzker Nanofabrication Facility of the Institute for Molecular Engineering at the University of Chicago, which receives support from Soft and Hybrid Nanotechnology Experimental (SHyNE) Resource (NSF ECCS-1542205), a node of the National Science Foundation’s National Nanotechnology Coordinated Infrastructure. This manuscript has been authored by Fermi Research Alliance, LLC under Contract No. DE-AC02-07CH11359 with the U.S. Department of Energy, Office of Science, Office of High Energy Physics, with support from its QuantISED program. We acknowledge support from the Samsung Advanced Institute of Technology Global Research Partnership.

\newpage
\renewcommand{\thesection}{\Alph{section}}
\renewcommand{\thefigure}{S\arabic{figure}}
\renewcommand{\thetable}{S\arabic{table}}
\renewcommand{\theequation}{S\arabic{equation}}



\onecolumngrid
\begin{center}
    \vspace{5ex}
    \centerline{\large \textbf{Stimulated emission of signal photons from dark matter waves}}
    \vspace{3ex}
    \centerline{\large \textbf{Supplemental Material}}
    \vspace{3ex}
    \normalsize Ankur Agrawal, Akash V. Dixit, Tanay Roy, Srivatsan Chakram, Kevin He,  Ravi K. Naik, \\ David I. Schuster, Aaron Chou
    \vspace{5ex}
\end{center}

\twocolumngrid

\section{Conceptual overview}
\begin{figure}[htbp]
    \centering
    \includegraphics[width=\columnwidth]{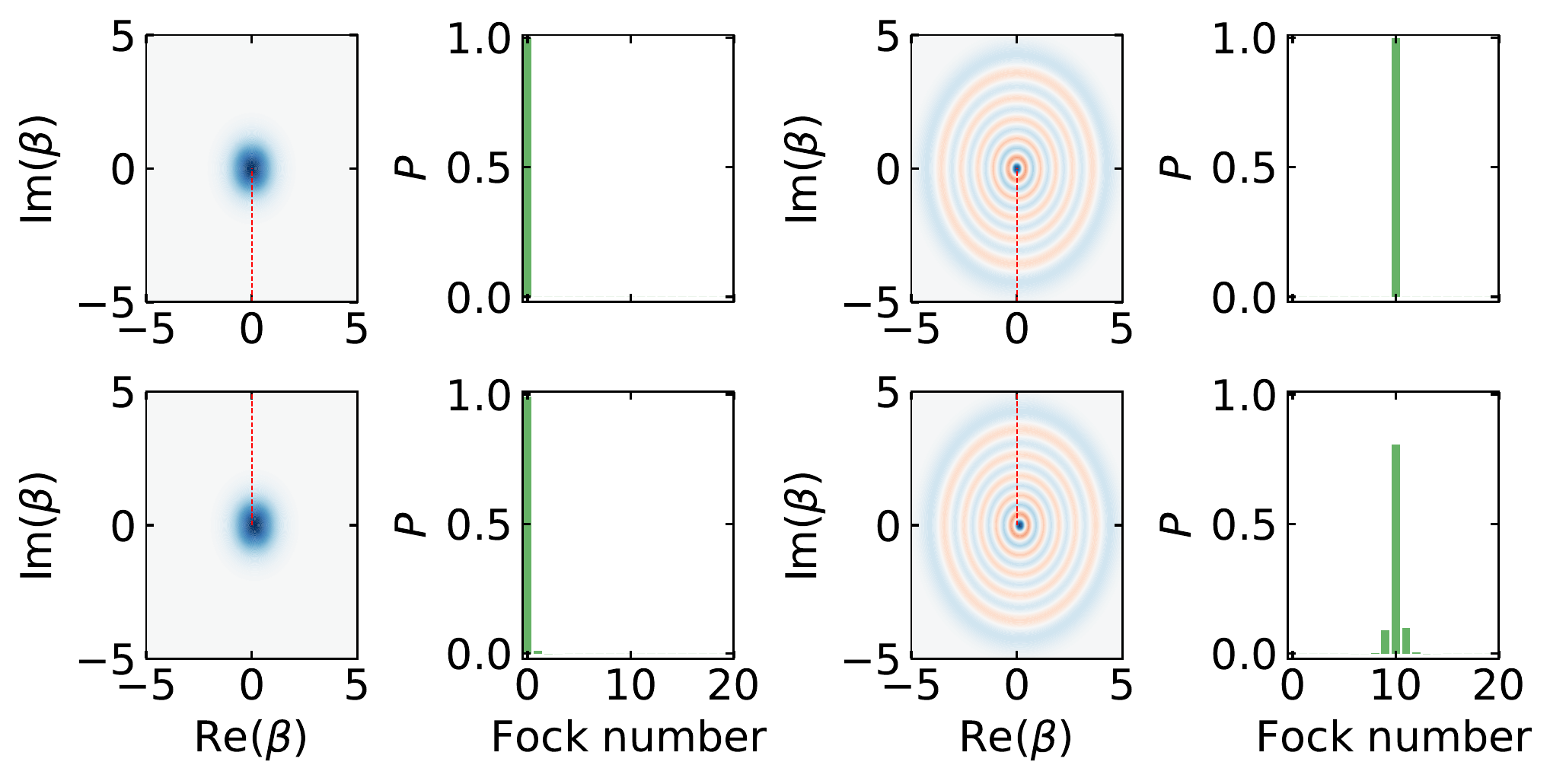}
     \caption{\textbf{Phase-space representation of the cavity state before and after the dark matter wave push.} (Left plots) Displacing the cavity initialized in $\ket{0}$ in an arbitrary direction by a small coherent push $\beta_{\rm coh} \ll 1$ results in a small probability $P_{0, 1} \propto |\mathcal{M}_{0, 1}|^{2}$ for creating a $\ket{1}$ component by spontaneous emission of a photon from the DM wave. The direction of displacement is determined by the instantaneous phase of the DM wave which is randomized every DM coherence time, but since the initial cavity state is azimuthally symmetric, a displacement in any direction gives the same probability $P_{01}$. The red dashed line is shows as a guide to locate the origin w.r.t to the center of the blob. (Right plots) The cavity is initialized in a $\ket{n}=10$ Fock state which also has an azimuthally symmetric Wigner distribution. Displacing this distribution in an arbitrary direction shifts some part of the distribution to larger radius and other parts to smaller radius. For example, the lower plots shows a displacement of $\beta_{\rm coh}=0.1$ in the positive $X=Re(\beta)$ direction. The shift to larger radius corresponds to stimulated emission to states with larger photon number, for example $\ket{11}$ while the shift to smaller radius corresponds to stimulated absorption to states with smaller photon number, for example $\ket{9}$. As shown in the histograms, the stimulated enhancement factors ($N_{Fock}+1$) and $N_{Fock}$ give probabilities which satisfy $P_{10, 11} = 11 \times P_{0, 1}$ and $P_{10, 9} = 10 \times P_{0, 1}$.}
     \label{fig:stim_em_concept}
\end{figure}

We use QuTip \cite{Johansson2012, Johansson2013} simulation tool to study the evolution of a cavity initialized in different Fock states under the action of a small displacement drive. As seen in Fig. \ref{fig:stim_em_concept}, the probability of observing a dark matter induced signal is significantly enhanced for a cavity prepared in large $\ket{n}$ Fock state.

\section{Dark matter conversion and scan rate}
The expected signal rate ($R_{s}$) from DM is proportional to the volume and the quality factor of the cavity. In DM experiments operating below $\mathrm{1\, GHz}$ ($h\nu \ll k_{B}T$), the background rate ($R_{b}$) is limited by the thermal occupation of the cavity such that the noise added by an amplifier in the readout process is sub-dominant. However, at frequencies in $\mathrm{5-30 \, GHz}$ range, the DM search faces major challenges as the signal-to-noise ratio (SNR) rapidly plummets down. 

The expected DM signal decreases at higher frequencies due to following reasons: (1) the cavity volume diminishes as the cavity dimensions shrink to maintain the resonance condition ($V\propto \nu^{-3}$) for a $\mathrm{TM_{0n0}}$ type mode, (2) the quality factor of a normal metal cavity degrades due to the anomalous skin effect \cite{Pippard1947, Reuter1948}. Moreover, the noise from quantum-limited readout process increases at higher frequencies because the power scales as  $h\nu \, (\frac{\nu}{Q_{\rm DM}})$, where $Q_{\rm DM}$ is the quality factor of a DM wave given by the escape velocity of DM from the galaxy. The first term in the expression represents the energy of one photon, which is determined by the standard quantum limit. This limit specifies that the minimum amount of noise added by an amplifier is equivalent to a mean photon number of $\bar{n}_{SQL}\geq 1$\cite{Caves1982, Yamamoto2008, Eichler2014, Tanay2015, Martina2019}. The signal frequency scan rate $R$ (in Hz/s), a key merit of haloscope type experiment, is proportional to $\propto R_{s}^{2}/R_{b}$ as shown in Eqn. \eqref{eq:scan_rate}. Quantum-enhanced techniques to improve signal rate and reduce noise are required to improve SNR and hence accelerate the search for the extremely weak dark matter signal.

The integration time ($\Delta t$) required for a background limited dark matter search is given by the time required to achieve $1\sigma$ sensitivity determined by Poisson counting statistics: $R_s \Delta t > \sqrt{R_b \Delta t}$, where $R_s = \bar{n}_{\mathrm{HP}}/\Delta t$ and background rate $R_b = \bar{n}_c^{b}/\Delta t$. The signal accumulation time for each Fock state preparation is given by $\tau$. For an optimal dark matter scanning experiment, we would want to start with a high-Q cavity ($Q_{\rm cav} > Q_{\rm DM}$) in the largest possible Fock state $\ket{n}$ such that $\tau$ (which must be smaller than the Fock state coherence time $Q_{cav}/(\nu * n)$) can be matched with the signal coherence time $Q_{\rm DM}/\nu$. Hence, the integration time at each frequency point scales as $R_{b}/R_{s}^{2}$ and, the scan rate is given by,
\begin{equation}
    R = \Delta \nu / \Delta t = (\frac{\nu}{Q_{\rm DM}})\frac{1}{N\, \tau} \propto R_{s}^{2}/R_{b},
\label{eq:scan_rate}
\end{equation}
such that $\Delta t$ is the sum of many iterations of duration $\tau$ and $N$ is the total number of iterations for each Fock state preparation. While the stimulated emission technique will generally enhance $R_s$ by a factor $(n+1)$, the coherence time of the prepared quantum state is also reduced by a factor $(n+1)$, thus also restricting the accumulation time $\tau$. The smaller coherence time then necessitates a faster readout rate which then generally causes the background rate $R_b$ to also increase by a factor $(n+1)$ for backgrounds which are associated with the readout procedure or due to a residual non-zero mode occupation number in the cavity. The overall improvement in $d\nu/dt$ due to Fock enhancement is then a single factor $\eta \, (n+1)$, where $\eta$ is the detection efficiency in the experiment.

Each measurement sequence consists of state preparation and validation ($\mathrm{2\, \mu s} + 3\times t_{m} = \mathrm{17\, \mu s}$), signal integration ($\mathrm{20\, \mu s}$) and 30 repeated measurements (30$\times t_{m} = \mathrm{150\, \mu s}$). There is an additional dead time in the experiment due to imperfect state preparation which constitutes $35\% \times \mathrm{17\, \mu s} \approx \mathrm{6\, \mu s}$ for $\ket{n=4}$ Fock state. The total search time at each Fock state is roughly $20, 000 \times \mathrm{193\, \mu s} = \mathrm{36.6 \, s}$ with a duty cycle of $\frac{20}{193} = 10\%$ ($\mathrm{3.6\, s}$ of integration). For the present apparatus, the duty cycle could be increased to $\sim 30\%$ if longer signal integration times are chosen at the expense of greater systematics due to Fock state decoherence.  Higher Q cavities would enable even larger integration times and duty cycles.

\section{Experimental setup}
\begin{figure}[htbp]
    \centering
    \includegraphics[width=\columnwidth]{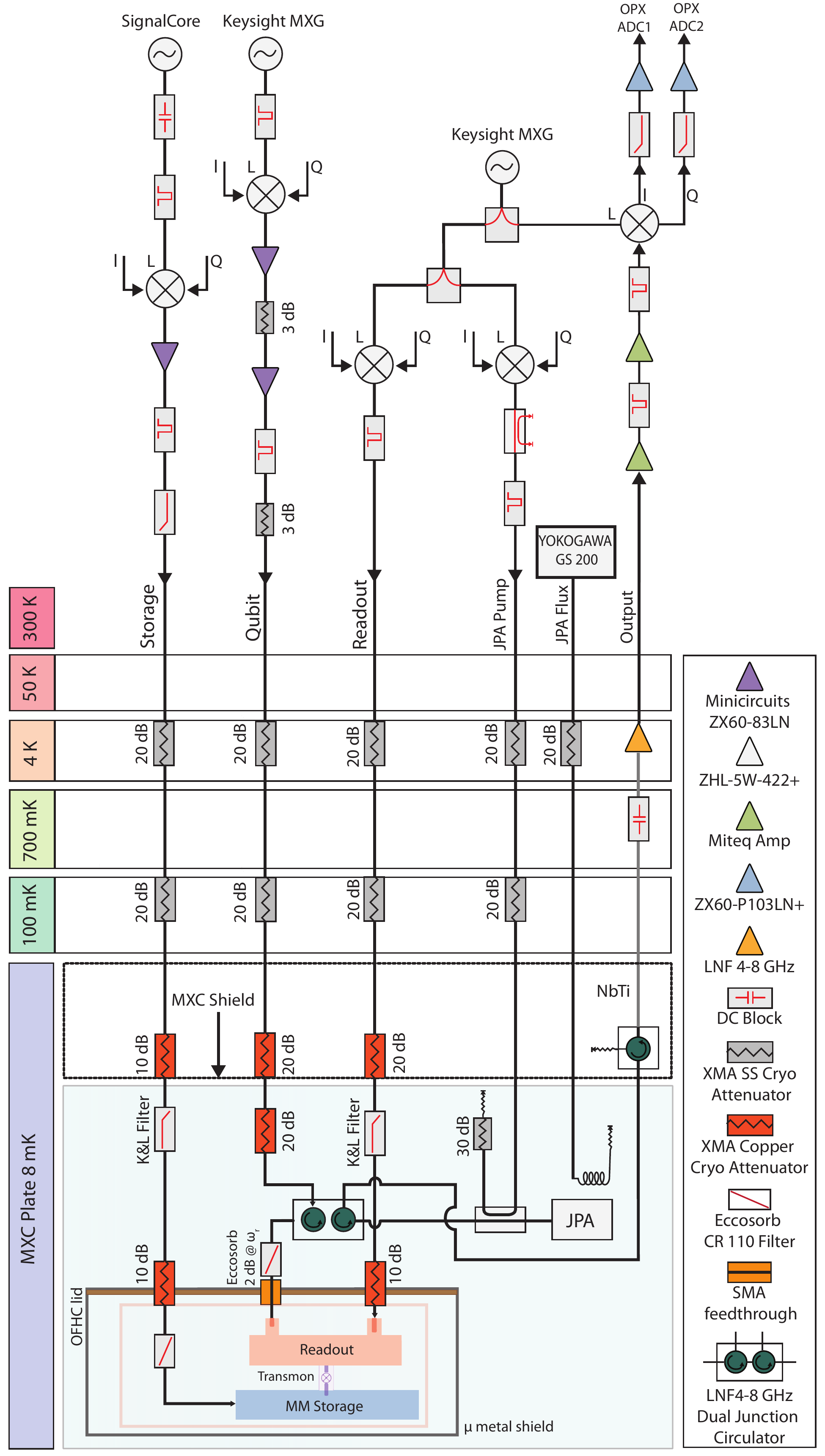}
    \caption[]{\textbf{Wiring diagram inside the dilution refrigerator and the room temperature measurement setup.} Quantum Machines (QM) OPX controller was used to generate the arbitrary wave forms (DACs) and digitize the incoming readout signal (ADCs). All the tones are up-converted using IQ modulation, with local oscillator (LO) and AWG synced  to an SRS F2575 RB source. Storage cavity is controlled via a direct port and qubit is controlled by injecting a drive into the strongly coupled port of the readout. Readout signal is injected into the weakly coupled port, and the signal is routed to the JPA using non reciprocal circulator elements. The amplified signal is routed to the HEMT for further amplification and the signal is then mixed down to $\mathrm{100\, MHz}$ IF, further amplified, and finally digitized. All the RF lines are heavily filtered with homemade eccosorb filters and attenuated to minimize stray radiation from entering the device.}
    \label{fig:wiring_diagram}
\end{figure}

The cavities and qubit are mounted to the base plate of a dilution fridge (Bluefors LD400) operating at $\mathrm{10\, mK}$. The device is housed in two layers of $\mu$-metal to shield from magnetic fields. Signals sent to the device are attenuated and thermalized at each temperature stage of the cryostat as shown in Fig. \ref{fig:wiring_diagram}. The field probing the readout resonator is injected via the weakly coupled port (shorter dipole stub antenna). Control pulse for the qubit are inserted through the strongly coupled readout port (longer dipole stub antenna). The storage cavity is driven through a direct port which is weakly coupled to the external microwave chain. Both control lines also contain an inline copper coated XMA attenuator that is threaded to the base plate. The signal from the readout resonator reflects off a Josephson parametric amplifier before being amplified by a cryogenic HEMT amplifier at the $\mathrm{4\, K}$ stage. The output is mixed down to $\mathrm{100 \, MHz}$ IF signal before being digitized. Cavity and transmon fabrication details can be found in Supplemental materials of Ref. \cite{Dixit2021, Chakram2022}.

\section{Fock state characterization}
We know that a Fock state has a definite photon number but no definite phase associated with it. Therefore, a simple qubit spectroscopy reveals as much as information as a Wigner tomogram would do. However, we use both these methods to compute and confirm the presence of Fock states in the cavity. Fig. \ref{fig:device_spec_wt} shows the qubit spectroscopy on the left and Wigner tomography of the resultant Fock states on the right. By fitting a Gaussian curve to the spectrum we estimate the fidelities to be $P_{0} = 95.2\pm0.3\%, P_{1} = 91.2\pm0.4\%, P_{2} = 87.3\pm0.5\%, P_{3} = 81.6\pm0.6\%, P_{4} = 63.6\pm0.7\%$. 

\begin{figure}[h]
    \centerline{
    \includegraphics[width=\columnwidth]{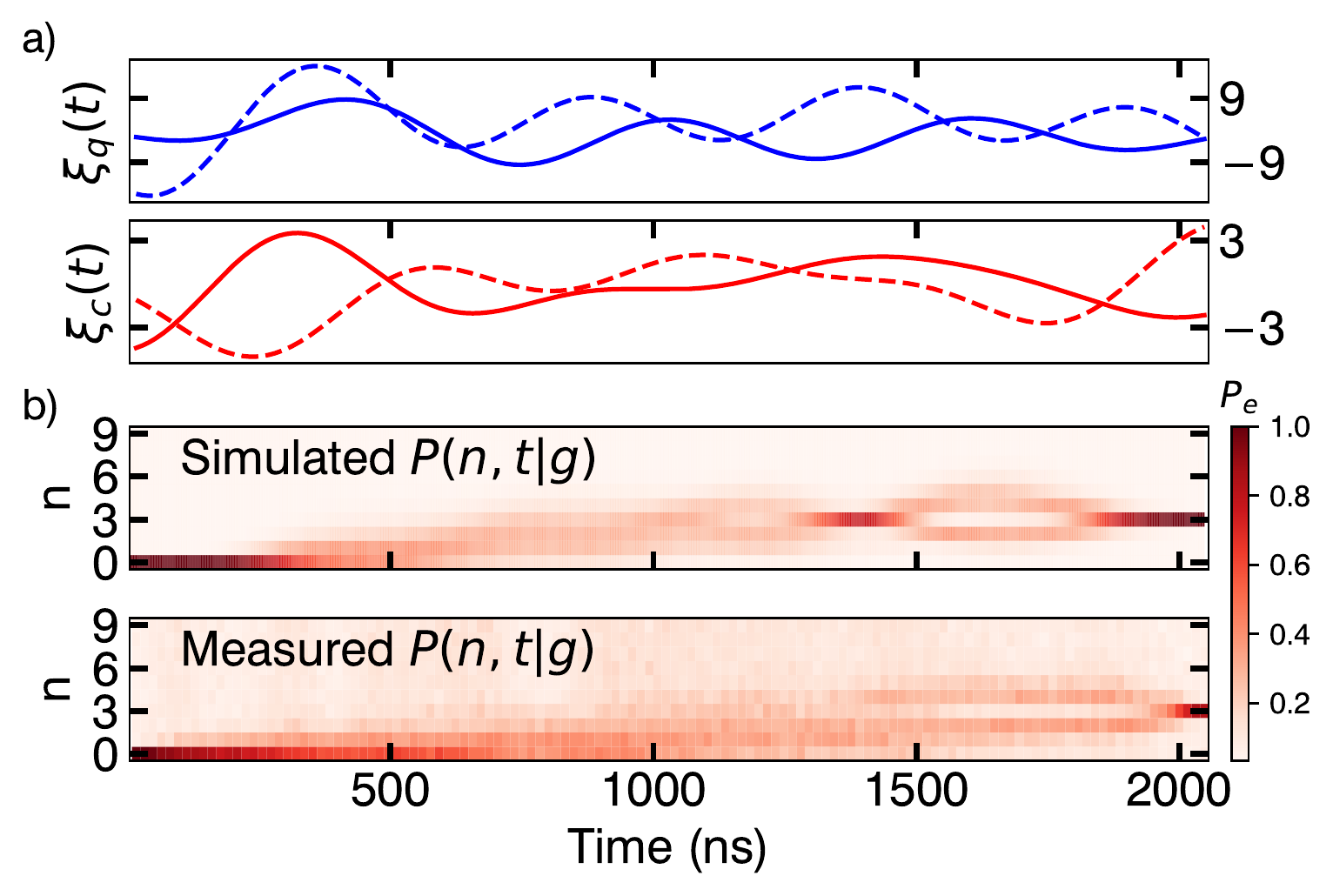}
    }
    \caption{\textbf{Fock state preparation in a cavity dispersively coupled to a transmon qubit}. \textbf{(a)} Optimized control drives to the qubit (top) and cavity (bottom) to transfer the qubit-cavity state from $\ket{g,0} \xrightarrow{} \ket{g, 3}$. The drive strengths are in units of $\mathrm{MHz}$. \textbf{(b)} Simulated trajectory of the cavity state using QuTip under the control drive. Measured cavity occupation probability showing all the levels explored during the evolution before converging to the desired state $\ket{g, 3}$ at the end. The experiment involves playing a fraction of the OCT pulses and performing qubit spectroscopy with a number resolved $\pi$-pulse. Each vertical slice corresponds to the cavity occupation probability in different number states as function of the duration of pulse. Note that the simulation does not include the presence of neighboring cavity modes as well as the measurement infidelity. Control fields for the qubit and cavity are generated by a room temperature field-programmable gate array (FPGA) controller, which also provides a low-latency feedback signal to actively monitor the state of the cavity.}
    \label{fig:GRAPE_traj}
\end{figure}

Moreover, we study the evolution of the cavity under the action of OCT pulses by tracking the cavity state as shown in Fig. \ref{fig:GRAPE_traj} (b). We perform a number resolved qubit spectroscopy at different points in time to map out the occupation probability of different Fock levels, in agreement with the simulated trajectory. 

In several recent works, it has been demonstrated that a single transmon is capable of preparing any quantum state in the cavity and perform universal control on it. The methods developed include - resonant swap of a qubit excitation to the cavity \cite{Wang2008}, SNAP gates \cite{Heeres2015}, Blockade \cite{Chakram2022}, and  GRadient Ascent Pulse Engineering (GRAPE) based optimal control (OCT) pulses \cite{Nelson2017, Heeres2017} and Echoed Conditional Displacement (ECD) control pulses \cite{Eickbusch2022} to solve the inversion problem of finding control fields to transfer a quantum system from state A to B. We use a GRAPE based method to generate a set of optimal control pulses \cite{Nelson2017, Heeres2017} to prepare non-classical states in an cavity. In the optimal control, we consider the full model of the time dependent Hamiltonian and generate the control pulses which maximizes the target state fidelity as has been previously demonstrated in nuclear magnetic resonance experiments \cite{Khaneja2005} as well as superconducting circuits \cite{Heeres2017}. The main advantage of this approach is that the duration of state preparation pulses can be as short as 1/$\chi$ and does not necessarily increase for higher Fock states \cite{Heeres2017, Eickbusch2022}. We use OCT pulses to successfully prepare cavity Fock states as shown in Fig. \ref{fig:device_spec_wt} (b).

We briefly investigated the SNAP protocol \cite{Heeres2015} to prepare Fock states but did not pursue further as it suffers from two issues limiting the maximum achievable fidelity. First, the number of constructed sequences scales as $(2n+1)$, requiring large number of gates, limiting operations which are feasible in the presence of decoherence. Second, the constructed model fails to account for the higher order Kerr non-linearity ($\mathcal{H}_{Kerr}/\hbar = K  (a^{\dagger})^{2} (a)^{2}$) in the Hamiltonian, which is non-negligible at higher photon occupation. This results in finite occupation probability at the other Fock states as well.  

\section{Calibration of number resolved \texorpdfstring{$\pi$ }pulse}
In order to successfully probe the presence of Fock states in the cavity, we need a very well calibrated narrow bandwidth $\pi$-pulse - amplitude as well as frequency. We perform an amplitude Rabi experiment with a $3\mu$s (4$\sigma$) long Gaussian pulse such that its spectral spread is smaller than the $\chi$-shift. After a coarse amplitude sweep, we perform a finer sweep centered at the initial guess and apply a set of 12 Gaussian pulses in succession to amplify any gate errors such as under/over rotation, frequency drift to get a better estimate of the $\pi$-pulse amplitude. For qubit transition frequency, we initialize the cavity in a particular Fock state $\ket{n}$ and perform a qubit Ramsey interferometry experiment. It consists of two $\pi/2$ pulses separated by a variable delay time. A Fast Fourier Transform (FFT) of the resultant time oscillations gives the shift in the qubit transition frequency which accounts for any higher order corrections as well. The computed transition frequencies and amplitude are used in the actual experiment to minimize the errors.

\section{Device  calibration}
\begin{figure}[hbt!]
    \centerline{
    \includegraphics[width=\columnwidth]{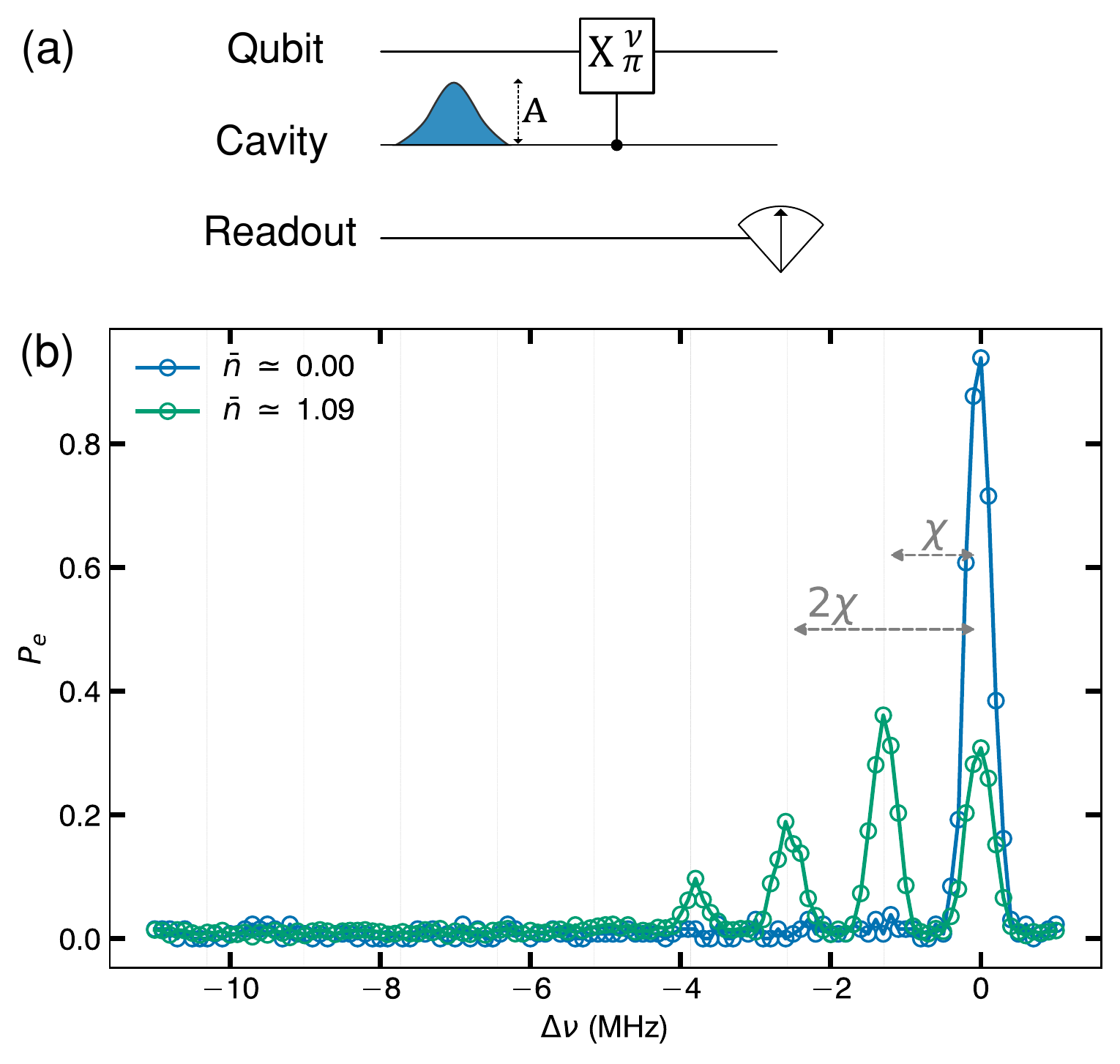}
    }
    \caption{\textbf{Qubit spectroscopy reveals cavity displacement} (a) Pulse sequence showing the calibration of cavity photon number by performing qubit spectroscopy. (b) The cavity is displaced using a variable weak coherent drive for a finite period of time. The resulting population of the cavity is determined by performing qubit spectroscopy with a resolved $\pi$-pulse. The cavity photon number dependent shift of the qubit transition frequency reveals the cavity population. By fitting the spectrum to a Poisson distribution we extract the weights of the cavity number states in the prepared coherent state and thus, the mean photon occupation.}
    \label{fig:alpha_cal}
\end{figure}

By applying a weak coherent tone at the storage cavity frequency, we induce a variable displacement $\alpha$ of the cavity state. We calibrate the number of photons injected into the storage cavity by varying the drive amplitude and performing qubit spectroscopy. By fitting the qubit spectrum as shown in Fig. \ref{fig:alpha_cal} to a Poisson distribution, we extract the cavity occupation, $\bar{n} = \mathrm{|\alpha|^{2}}$. Similarly, the qubit drive strength is calibrated by performing qubit Rabi experiments with varying AWG amplitude and pulse length. The measured drive strength and the AWG amplitude mapping is used to send the correct OCT pulses to the device. See Supp. section in \cite{Dixit2021} and \cite{Chakram2021} for more details on device parameter calibrations.

\section {Hidden Markov model analysis}
We adopt the Hidden Markov Model (HMM) approach presented in \cite{Dixit2021} to consider all the possible errors that may occur during the measurement process and alter the state of the cavity, qubit, and readout. The cavity and qubit states are treated as hidden variables that emit readout signals. The Markov chain is modeled by the transition matrix ($T$) (Eqn. \eqref{eqn:T_matrix}) which describes the evolution of the joint cavity-qubit hidden state $s \in [\ket{n'g}, \ket{n'e}, \ket{ng}, \ket{ne}]$ and the emission matrix ($E$) (Eqn. \eqref{eqn:E_matrix}) that calculates the probability of a readout signal $R \in$ [$\mathcal{G}$, $\mathcal{E}$] given a certain hidden state. It's important to note that with a number-resolved $\pi$-pulse centered at the $n$-shifted peak, we can only determine whether the cavity is in the $\ket{n}$ Fock state or not ($\ket{n'\neq n}$).

The transition and emission matrix were first introduced in \cite{Zheng_2016} and implemented in \cite{Dixit2021}. The change in qubit state probability is not affected by the cavity state. However, the repeated measurement of the cavity state with a number selective $\pi$-pulse introduces an additional channel through which the cavity loses its excitation. It is called demolition probability ($p_{d}$), which signifies the non-QNDness of the projective measurement of the cavity state with the qubit. In simple terms, a single qubit measurement is $(1-p_{D})\%$ QND. We followed the protocol described in Ref.\cite{Sun2014} for repeated parity measurement and replaced it with number selective $\pi$-pulse. Table \ref{table:qndness} and Fig. \ref{fig:qndness_storage} show the measured values for different Fock states. The elements of the emission matrix represent the readout fidelities for the ground and excited states of the qubit, which are influenced by noise from the first stage JPA. More information on the experimental protocols used can be found in the supplemental material.

We used the backward algorithm (Eqn. \eqref{eqn:backward}) \cite{Zheng_2016, Hann2018} to calculate the probability ($P(n_0)$) that the cavity state was in the $\ket{n+1}$ Fock state after the injection of a synthetic DM signal. The algorithm takes a set of $N+1$ measured readout signals ($R_{0}, R_{1}, ..., R_{N}$) as input, where $n_{0} = n+1$ represents the target Fock state and $n_{0}'$ can take on values in the range [$0, n, n+2, ...$].

\begin{equation}
\begin{aligned}
    P(n_0) = \sum_{s_0 \in [\ket{n_0,g}, \ket{n_0,e}]} \sum_{s_1} ... \sum_{s_N} & E_{s_0,R_0} T_{s_0,s_1} E_{s_1,R_1} \\
    \cdots & T_{s_{N-1},s_N} E_{s_N,R_N}
    \label{eqn:backward}
\end{aligned}
\end{equation}

In the reconstruction process, all possible scenarios are taken into account. For instance, a readout measurement of $\mathcal{G}$ followed by $\mathcal{E}$ could happen due to the successful detection of the cavity in the $\ket{n+1}$ state with a probability of $P_{n_{0}n_{0}}P_{gg}F_{e\mathcal{E}}/2$. On the other hand, it could also be caused by either a qubit heating event ($P_{n_{0}'n_{0}'}P_{ge}F_{e\mathcal{E}}/2$) or a readout error ($P_{n_{0}'n_{0}'}P_{gg}F_{g\mathcal{E}}/2$). The data in Fig. \ref{fig:pulse_meas}(b) illustrates the results of the measurement of the readout signals and the reconstructed initial probabilities of the cavity state. The panels on the left show instances when there was no emission event, while the panels on the right depict cases where a positive emission event took place. The cavity state shifted from $\ket{1}$ to $\ket{2}$ as a result of a variable displacement drive $\mathcal{D}(\alpha)$, and the change was accurately reflected in the reconstructed probability.

The reconstructed cavity state probabilities undergo a likelihood ratio test $(\lambda =\frac{P(n_0=n+1)}{P(n_0\neq n+1)})$ to determine if the cavity state changed or not. A positive detection of a photon signal $\ket{n} \rightarrow \ket{n+1}$ is declared when the threshold $\lambda > \lambda_{\mathrm{thresh}}$. The false positive rate is limited to less than $\frac{1}{\lambda_{\mathrm{thresh}}+1}$ as a result of this procedure. A higher detection threshold can be achieved by increasing the number of repeated measurements, but at the cost of decreased efficiency which is linear in the number of measurements. Hence, $\lambda_{\mathrm{thresh}}$ is chosen to strike a balance between the detection efficiency and the false positive rate, keeping it below the observed physical photon background. This will be discussed further in the next section.

\section{Elements of hidden Markov model}
The hidden Markov model relies on independent measurements of the probabilities contained in the transition and emission matrices. The elements of these matrices depend on the parameters of the experiment and the device, including the lifetimes of the qubit and cavity, qubit spurious population, and readout fidelities.

\subsection*{Transmission matrix elements}
\begin{figure}[hbt!]
    \centerline{
    \includegraphics[height=0.5\columnwidth]{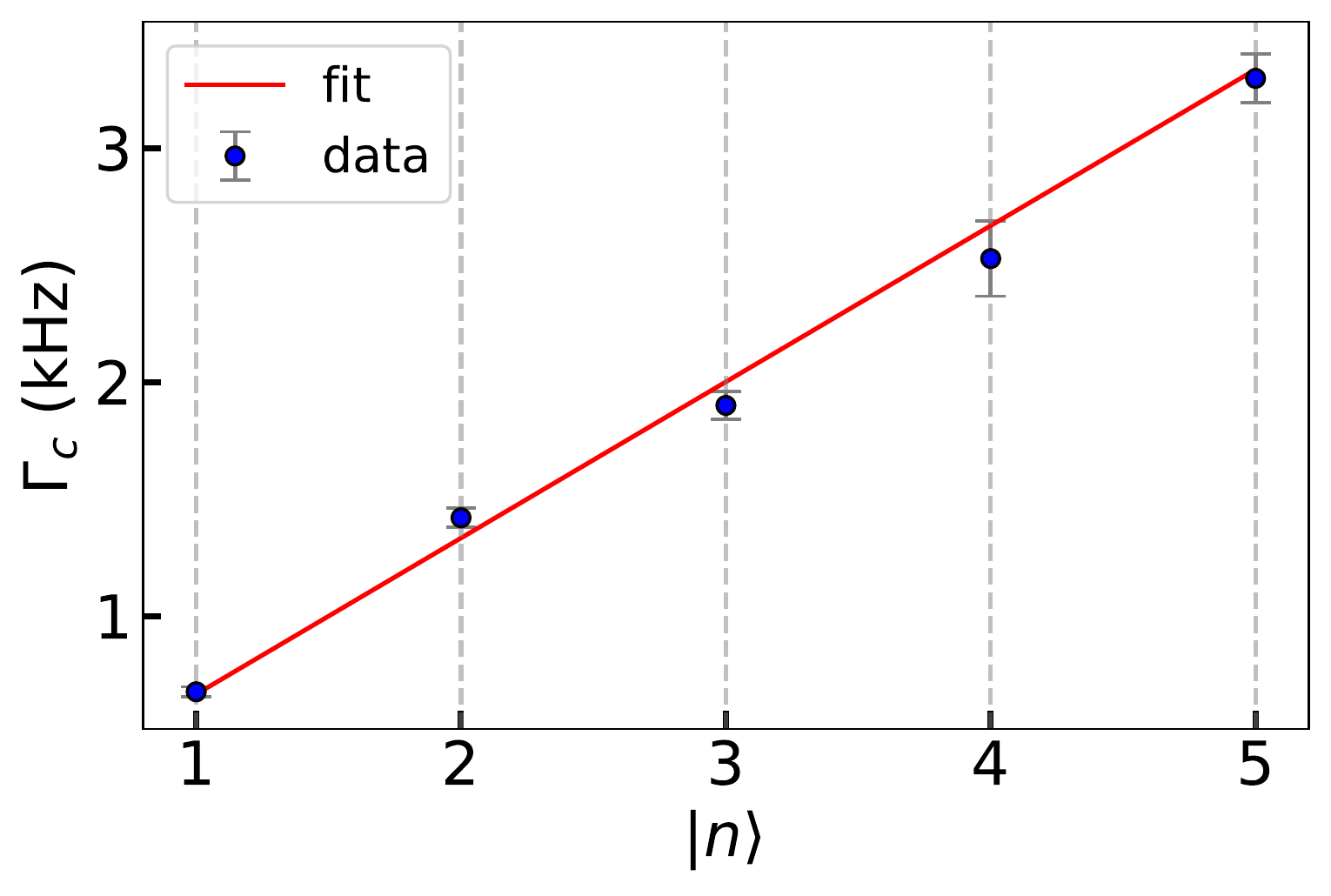}
    }
    \caption{\textbf{Decay rate of Fock states.} Measured lifetime of the different Fock states prepared in the cavity using GRAPE pulses. The decay rate is inversely proportional to the lifetime ($\Gamma = 1/T_{1}$). The curve shows enhancement in the decay rate as a function of the Fock state $\ket{n}$. The linear fit to the data fits well as predicted by \cite{Lu1989}, where $T^{n}_{1} = T_{1}/n$.}
     \label{fig:fock_state_decay}
\end{figure}

\begin{figure}[hbt!]
    \centerline{
    \includegraphics[height=0.5\columnwidth]{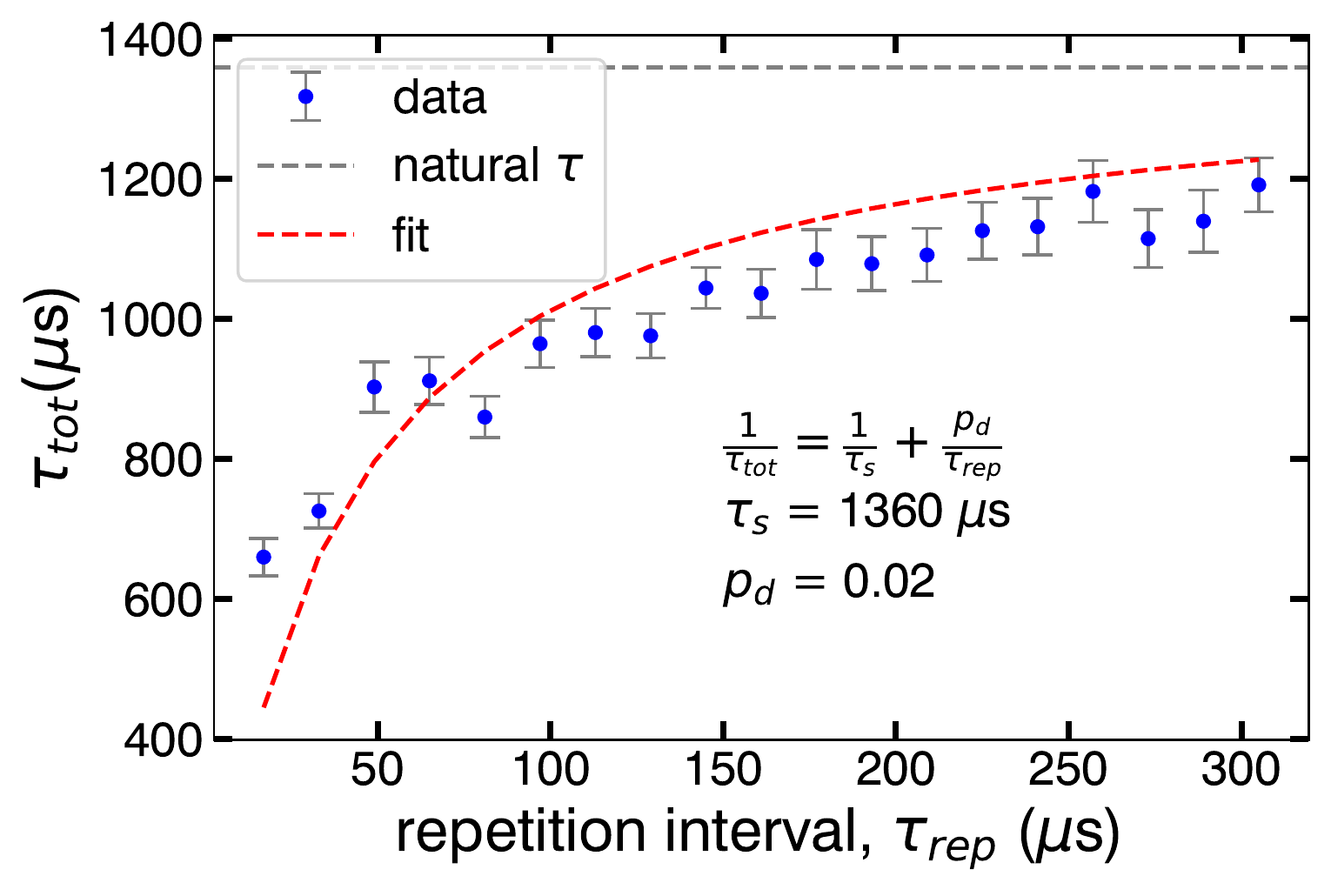}
    }
    \caption{\textbf{QNDness of storage cavity measurement.} Storage cavity $T_{1}$ measurements were performed with repeated number resolved qubit measurements interleaved during the delay time with a variable repetition interval time $\tau_{\rm rep}$. The extracted total decay time was fit to a model $1/\tau_{\rm tot} = 1/\tau_{s} + p_{d}/\tau_{\rm rep}$. From the fit (red line), we infer a demolition probability per readout of $p_{d}=2.6\%$ corresponding to a QNDness of  97.4\%, which is a bit lower than reported for a parity protocol \cite{Sun2014}. The natural decay time of the storage $\tau_{s}=\mathrm{1360\, \mu s}$ is indicated by a dashed grey line.
    }
     \label{fig:qndness_storage}
\end{figure}

The transition matrix captures the possible qubit (cavity) state changes. Qubit relaxation $\ket{e} \rightarrow \ket{g}$ occurs with a probability $P_{eg}^{\downarrow} = 1 - e^{-t_m/T_1^q}$. The probability of spontaneous heating $\ket{g} \rightarrow \ket{e}$ of the qubit towards its steady state population is given by $P_{ge}^{\uparrow} = \bar{n}_q [1 - e^{-t_m/T_1^q}]$. Unlike a two-level system such as a qubit, the cavity state may change from $\ket{n} \rightarrow \ket{n'}$ via either decay ($\ket{n} \rightarrow \ket{n-1}$) or excitation ($\ket{n} \rightarrow \ket{n+1}$) with probabilities $P_{n, n-1} = 1 - e^{-t_m/T_1^n}$ or $P_{n, n+1} = \bar{n}_c [1 - e^{-t_m/T_1^n}]$ respectively. The lifetime of the Fock state is modified due to the enhanced decay \cite{Wang2008} ($T_1^{n} = T_1^{s}/n$) as compared to the bare lifetime of a coherent state (see Supp Fig. \ref{fig:fock_state_decay}). Yet another possible source of change in the cavity state is the repeated qubit measurement itself. In most cases, we assume the interaction between the qubit and cavity to be QND, i.e., the measurement of cavity state by qubit does not perturb the state. However, we measured the QNDness of repeated qubit measurement for $\ket{n}=1$ Fock state to be $97.4\%$ which corresponds to a demolition probability $p_{d}$ of $2.6\%$ per measurement (see Fig. \ref{fig:qndness_storage}). Hence, we add this term in the transition matrix. All these probabilities are computed using independently measured qubit, cavity and readout parameters described below.
\begin{equation}
    T =
    \begin{blockarray}{ccccc}
    \ket{n'g} & \ket{n'e} & \ket{ng} & \ket{ne} \\
    \begin{block}{[cccc]c}
    P_{n'n'}P_{gg} & P_{n'n'}P_{ge} & P_{n'n}P_{ge} & P_{n'n}P_{gg} & \hspace{3pt} \ket{n'g} \\
    P_{n'n'}P_{eg} & P_{n'n'}P_{ee} & P_{n'n}P_{eg} & P_{n'n}P_{ee} & \hspace{3pt} \ket{n'e}\\
    P_{nn'}P_{gg} & P_{nn'}P_{ge} & P_{nn}P_{ge} & P_{nn}P_{gg} & \hspace{3pt} \ket{ng}\\
    P_{nn'}P_{eg} & P_{nn'}P_{ee} & P_{nn}P_{ee} & P_{nn}P_{eg} & \hspace{3pt} \ket{ne}\\
    \end{block}
    \end{blockarray}
    \label{eqn:T_matrix}
\end{equation}

\begin{figure}[hbt!]
    \centerline{
    \includegraphics[width=\columnwidth]{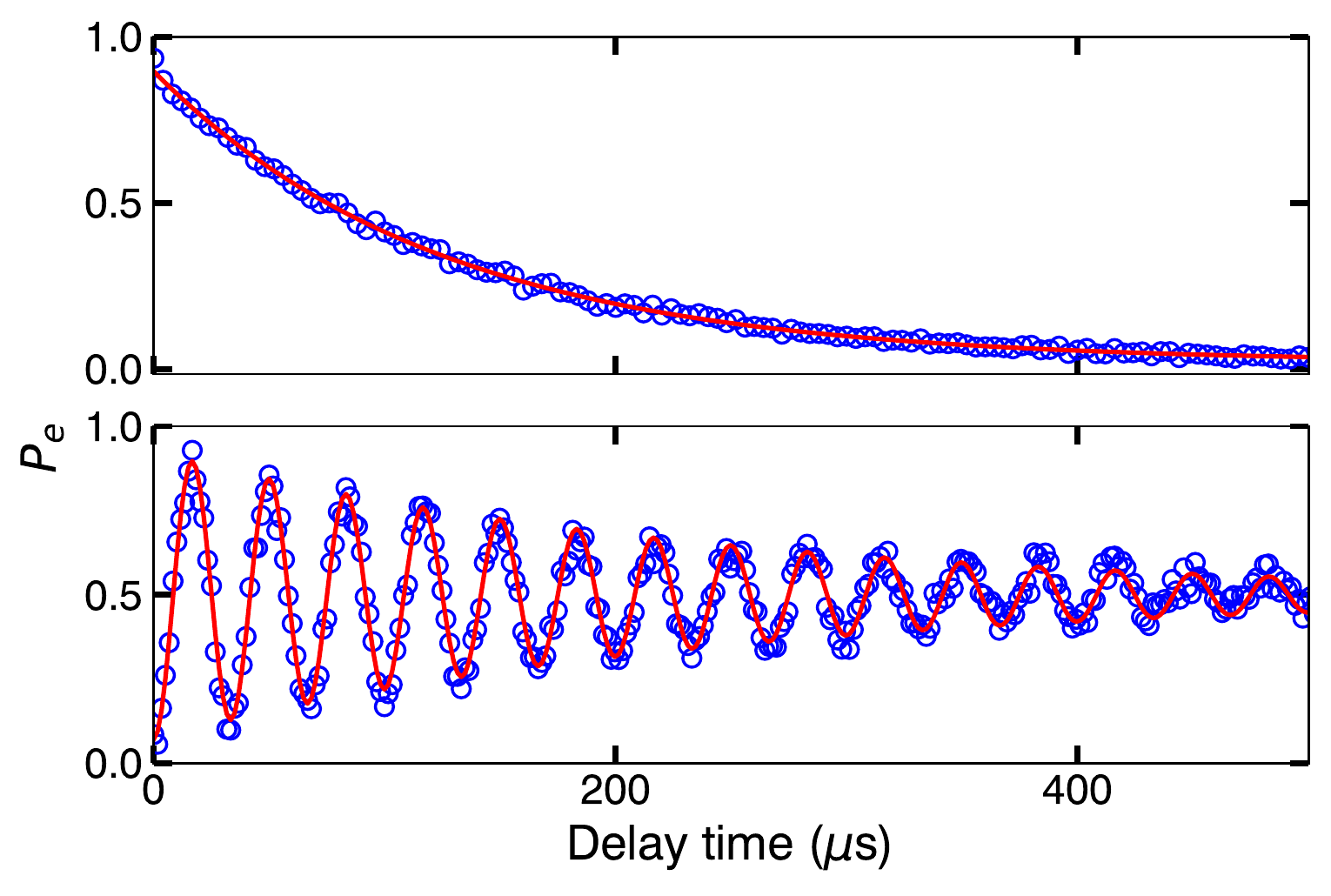}
    }
    \caption{\textbf{Qubit lifetime and dephasing time measurement.} (Top) $T_{1}$ measurement by sending a $\pi$-pulse to excite the transmon to $\ket{e}$ state and monitor its decay as a function of variable delay time. By fitting an exponential function to the qubit excitation probability $P_{e}$, we extract the $T_1^q = \mathrm{115\pm10\, \mu s}$. (Bottom) $T_{2}$ measurement with a Ramsey experiment. The sequence consists of two $\pi/2$-pulse separated by a variable delay time. The envelope of the measured oscillations informs the $T_{2}^{e}= \mathrm{236\pm6\, \mu s}$ and its frequency provides us the detuning between the drive and the transmon resonance frequency. In this case, we intentionally introduced a 60 kHz synthetic detuning.}
     \label{fig:qubit_coh}
\end{figure}

The lifetime of the qubit is determined by applying a $\pi$ pulse and waiting for a variable time before measuring the population as shown in Fig. \ref{fig:qubit_coh}. We map out the qubit population as a function of the delay time, fit it with an exponential characterizing the Poissonian nature of the decay process, and obtain $T_1^q = \mathrm{115\pm10\, \mu s}$.

The dephasing time of the qubit is measured by a Ramsey interferometry experiment with a $\pi/2$ pulse, variable delay, and a final $\pi/2$ with its phase advanced by $\omega_r t$ where $\omega_r$ is the Ramsey frequency. The phase advancement is implemented in the software. During the variable delay period, a series of $\pi$ pulses are applied to perform spin echos and reduce sensitivity to low frequency noise. We observe a dephasing time of $T_2^q = \mathrm{160\pm10\, \mu s}$ which extends to $T_{2}^{e}= \mathrm{236\pm6\, \mu s}$ with a single echo sequence.

\begin{figure}[hbt!]
    \centerline{
    \includegraphics[width=\columnwidth]{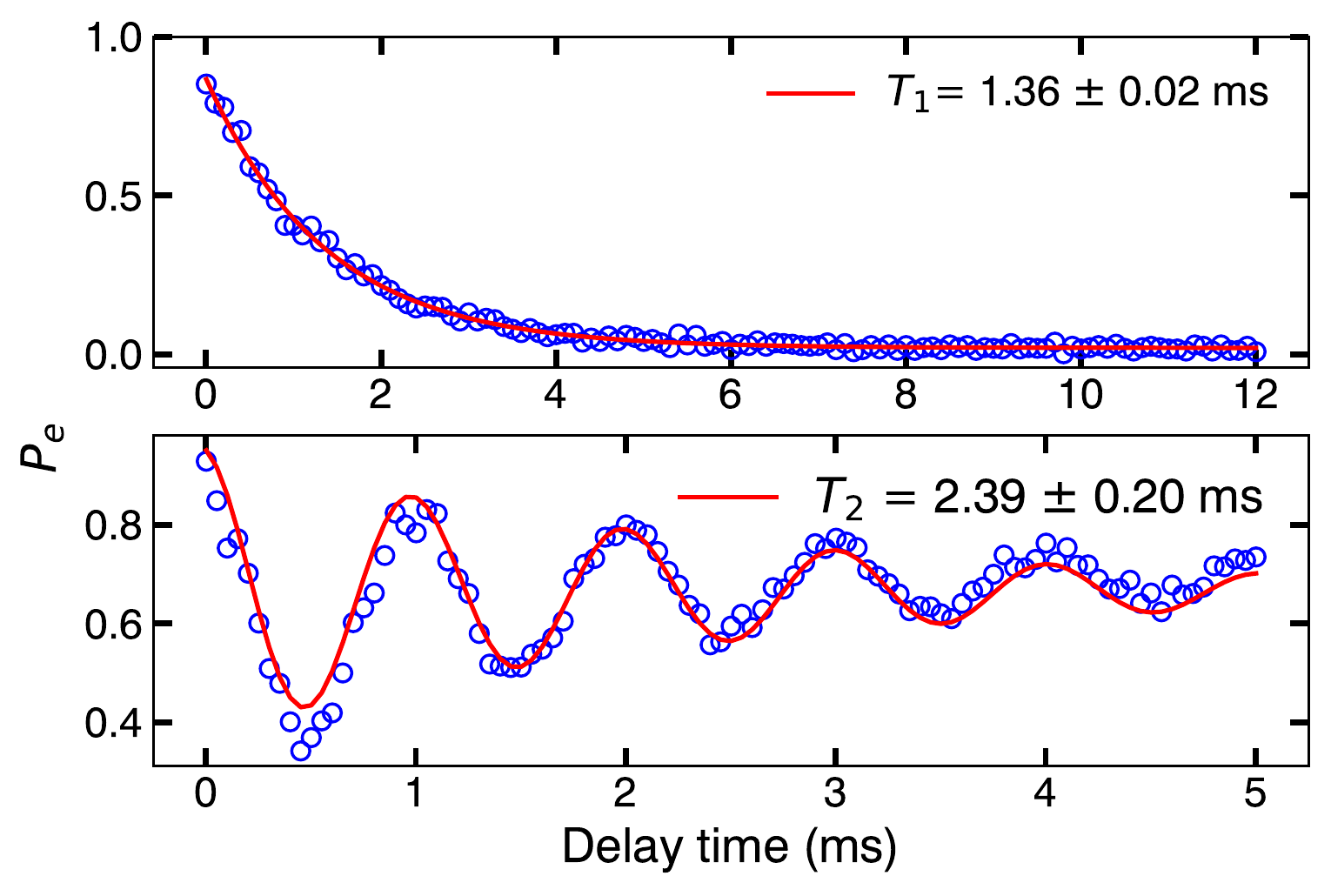}\
    }
    \caption{\textbf{Storage cavity lifetime and dephasing time from $\mathrm{T_1}$ and Ramsey measurements.} The long lived storage cavity mode is ideal for holding a signal photon induced by the dark matter while a series of repeated photon counting measurements is performed.}
    \label{fig:cavity_coherence}
\end{figure}

The storage cavity lifetime is calibrated by performing a cavity $T_1$ experiment. We use the OCT pulse to prepare the cavity in a $\ket{n}=1$ Fock state, wait for a variable delay time and probe the cavity state at the end by Rabi driving the qubit with a resolved $\pi$-pulse. The resultant cavity population is fitted to an exponential to obtain $T_1^s = \mathrm{1.36\pm0.02\,  ms}$ as shown in Fig. \ref{fig:cavity_coherence}. To measure the cavity dephasing time, the cavity is initialized in a superposition state $ (a \ket{0} + b \ket{1})$ by applying a weak coherent drive ($\alpha$) such that approximately only the first two photon states $\ket{n}$ are populated. The contrast in the signal will be determined by the relative amplitude ($\frac{a^{2}}{a^{2}+b^{2}}$) without any loss of information. After a variable delay time, an identical displacement drive with its phase advanced by $\omega_{r}\, t$ is applied before probing the $\ket{n}=0$ with a resolved $\pi$-pulse on the qubit. The oscillations are fitted to obtain a dephasing time of $T_2^s = \mathrm{2.39\pm0.02\,  ms}$.

The qubit spurious population is determined by measuring the relative populations of its ground and excited states \cite{Jin2015}. This is done by utilizing the $f$-level of the transmon. Two Rabi experiments are conducted swapping population between the $\ket{e}$ and $\ket{f}$ levels. First, we apply a $\pi_{ge}$ pulse to invert the qubit population followed by the $\ket{e}-\ket{f}$ Rabi experiment. Second, no $\pi_{ge}$ pulse is applied before the $ef$ Rabi oscillation. The ratio of the amplitudes of the oscillations gives us the ratio of the populations of the excited and ground state. Assuming that $P(g) + P(e) = 1$ and measuring $\frac{P(e)}{P(g)}=0.02$, corresponds to an effective qubit temperature of $\mathrm{54\, mK}$.

In a dispersive interaction, we assume that the measurement of cavity state via a parity or number resolved $\pi$-pulse does not perturbed the state i.e. it is a quantum non-demolition (QND) measurement or in other words, does not induce additional relaxation in the cavity mode. However, a recent study has shown that parity measurements, while highly QND, can induce a small amount of additional relaxation \cite{Sun2014}. Hence, in order to estimate the same, but, in the context of number resolved qubit measurement, we follow the method described in \cite{Sun2014}, where we perform a cavity $T_{1}$ experiment interleaved with varying number of repeated number resolved qubit measurements during the delay time. In that experiment, the total relaxation rate was modeled as a combination of the bare storage lifetime $\tau_{s}$ and a demolition probability $p_{d}$ associated with each qubit measurement. In Fig. \ref{fig:qndness_storage}, we show the extracted total decay time ($\tau_{\rm tot}$) and demolition probability $p_{d}=2 \pm 0.02\%$ when the cavity is prepared in $\ket{n}=1$ Fock state. In other words, a single number resolved qubit measurement is $98\%$ QND. Unfortunately, for $\ket{n}=2$, $p_{d}=4\pm0.04\%$ gets worse, which is not a good sign but probably explains why the detection efficiency in Fig. \ref{fig:det_char_stimem} falls off at higher Fock states. This acts like an additional source of loss to the cavity mode but only when the resolved $\pi$-pulse is on resonance with the shifted qubit frequency.

\begin{table}[hbt!]
\begin{center}
\begin{tabular}{c | c| c| c }
\hline
 $\ket{n}$ &$\tau_{s}(\mu s)$ & $p_{d}$ & $\sigma_{p_{d}}$ \\
 \hline
 $\ket{1}$ & 1360 & 0.026 & 0.002 \\
 $\ket{2}$ & 660 & 0.040 & 0.004 \\
 $\ket{3}$ & 527 & 0.10 & 0.05 \\
 $\ket{4}$ & 319 & 0.074 & 0.012 \\
\hline
\end{tabular}
\caption{\textbf{Demolition probability}. Measured lifetime of the different Fock states and their fitted demolition probability with error bars. $p_{d}$ can be approximated with a linear dependence on $n$.}
\label{table:qndness}
\end{center}
\end{table}

\subsection{Emission matrix elements}
\begin{figure}[hbt!]
    \centerline{
    \includegraphics[width=\columnwidth]{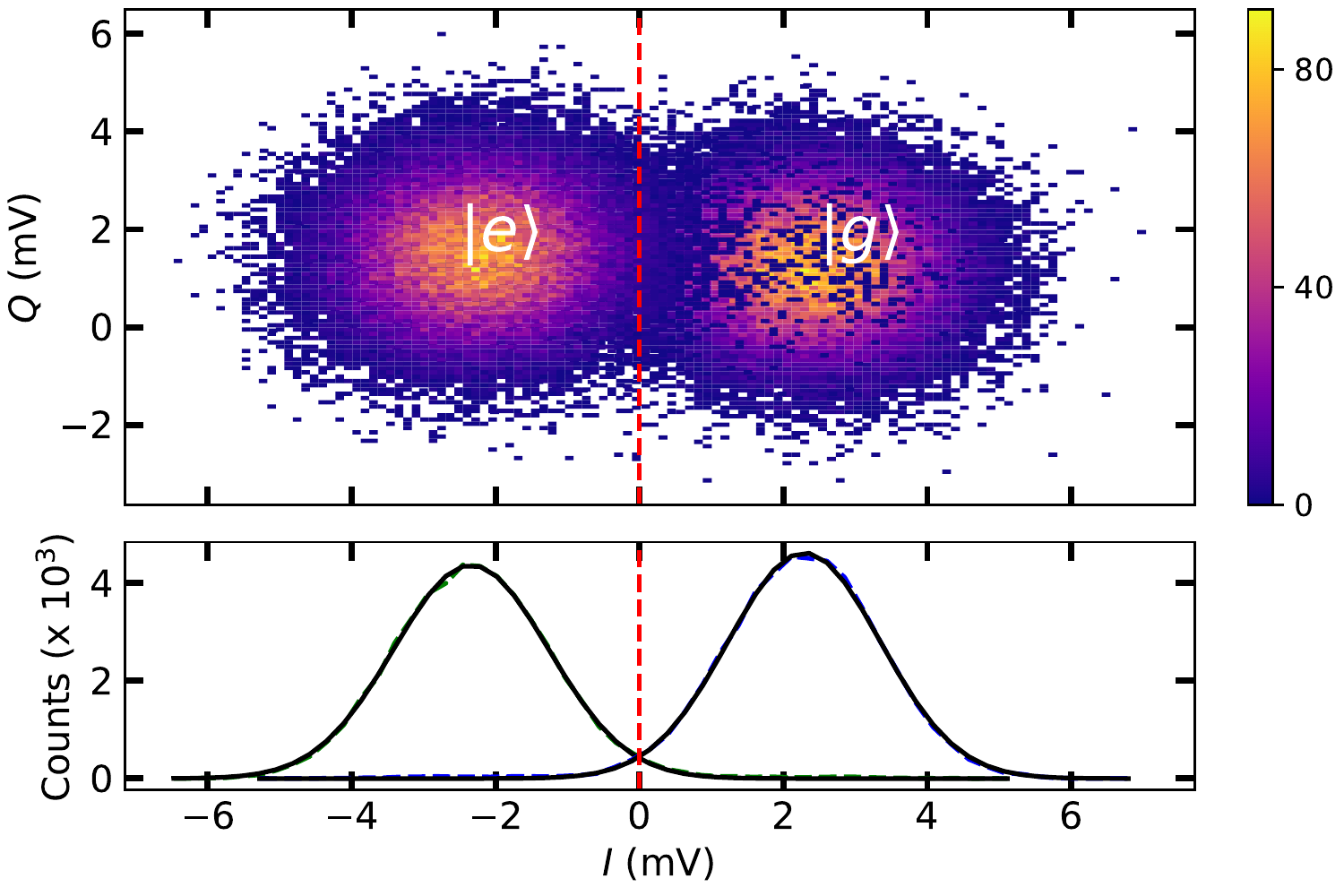}
    }
    \caption{\textbf{QND readout of the transmon state.} (Top) shows the two quadrature values of the down-converted readout signal for qubit prepared in $\ket{g}$ and $\ket{e}$ state. Each state preparation contains 20,000 single shot points. (Bottom) By fitting a sum of two Gaussian we can estimate the overlap region and assign a single-shot readout fidelity as $\mathcal{F} = 97\%$. The red dashed line shows an optimal value of the threshold for tagging a qubit state based on a single shot readout.}
    \label{fig:readout_hist_map}
\end{figure}

In order to characterize the emission matrix it is necessary to measure the readout infidelity of a particular transmon state. We consider only two possible transmon states ($\ket{g}, \ket{e}$) in this case as the number resolved $\pi$-pulses have very narrow spectral width ($\sigma_{\nu} \ll \alpha_{q}$). Each state is prepared 20,000 times and the resultant quadrature values are digitized to assign a voltage in the ($I-Q$) space. The phase of the readout pulse is pre-calibrated to align the signal along the $I$ axis. The histogram corresponding to each state is fitted with a sum of two Gaussian functions to estimate the overlap region and calculate the readout fidelity, $\mathcal{F}=97\%$ (Fig. \ref{fig:readout_hist_map}). A discriminator value (red dashed line) is used to assign each readout signal either $\mathcal{G}$ or $\mathcal{E}$ in real-time.

\begin{equation}
    E = \frac{1}{2} \hspace{3pt}
    \begin{blockarray}{ccc}
        \mathcal{G} & \mathcal{E} \\
        \begin{block}{[cc]c}
        F_{g\mathcal{G}} & F_{g\mathcal{E}} & \hspace{3pt} \ket{n'g}\\
        F_{e\mathcal{G}} & F_{e\mathcal{E}} & \hspace{3pt} \ket{n'e}\\
        F_{g\mathcal{G}} & F_{g\mathcal{E}} & \hspace{3pt} \ket{ng}\\
        F_{e\mathcal{G}} & F_{e\mathcal{E}} & \hspace{3pt} \ket{ne}\\
        \end{block}
    \end{blockarray}
    \label{eqn:E_matrix}
\end{equation}

Readout errors are due to voltage excursions from amplifier noise or spurious qubit transitions. The emission matrix should only contain readout errors that occur due to voltage fluctuations. Errors due to qubit transitions during the readout window are accounted for in the transition matrix. To disentangle the two contributions, we subtract the readout errors caused by the spontaneous heating and decay of the qubit to obtain $F_{g\mathcal{G}} = \mathrm{97.5\pm 1\%}$ and $F_{e\mathcal{E}} = \mathrm{96.8\pm 1\%}$.

\begin{table}[hbt!]
\begin{center}
\begin{tabular}{ l l }
\hline
 Device Parameter & Value \\
 \hline
 Qubit frequency & $\omega_q = 2 \pi \times \mathrm{4.961\, GHz}$ \\
 Qubit anharmonicity & $\alpha_q = - 2 \pi \times \mathrm{143.2\, MHz}$ \\
 Qubit decay time & $T_1^q = \mathrm{115 \pm10 \, \mu s}$ \\ 
 Qubit dephasing time & $T_2^q = \mathrm{160 \pm 10\, \mu s}$ \\ 
 Qubit echo time & $T_2^e = \mathrm{236\pm6\, \mu s}$ \\ 
 Qubit residual occupation & $\bar{n}_q = \mathrm{2\pm1} \times 10^{-2}$ \\
 Storage frequency & $\omega_s = 2 \pi \times \mathrm{5.965\, GHz}$ \\
 Storage decay time & $T_1^s = \mathrm{1360\pm23\, \mu s}$ \\
 Storage dephasing time & $T_2^s = \mathrm{2390\pm286\, \mu s}$ \\
 Storage-Qubit Stark shift & $\chi =  - 2 \pi \times \mathrm{1.285\, MHz}$ \\
 Storage residual occupation &  $\bar{n}_c = \mathrm{6.3 \pm3} \times 10^{-3}$ \\
 Readout frequency & $\omega_r = 2 \pi \times \mathrm{7.790\, GHz}$ \\
 Readout $\ket{e}$ shift & $2\chi_r^{e} =  - 2 \pi \times \mathrm{1.53\, MHz}$ \\
 Readout fidelity ($\ket{g}$) & $F_{g\mathcal{G}} = \mathrm{97.5\pm1\%}$ \\
 Readout fidelity ($\ket{e}$) & $F_{e\mathcal{E}} = \mathrm{96.8\pm1\%}$ \\
\hline

\end{tabular}
\caption{\textbf{Device parameters}. Measured qubit, storage, and readout cavity parameters. These independently measured values are necessary to determine for the transition and emission matrices. This enables the hidden Markov model to capture the behavior of the system during the measurement sequence.}
\label{table:device_params}
\end{center}
\end{table}

\section{Detector characterization}
To characterize the detector, the cavity population is varied by applying a weak drive and the cavity photon number is counted using the technique described in the main text. In order to extract the efficiency ($\eta$) and false positive probability ($\delta$) of the detector, the relationship between injected photon population ($\bar{n}_\mathrm{inj}$) and measured photon population ($\bar{n}_\mathrm{meas}$) is fit to $\bar{n}_\mathrm{meas} = \eta \times \bar{n}_\mathrm{inj} + \delta$.

\subsection*{Detector efficiency}
\begin{figure}[hbt]
    \centerline{
    \includegraphics[width=\columnwidth]{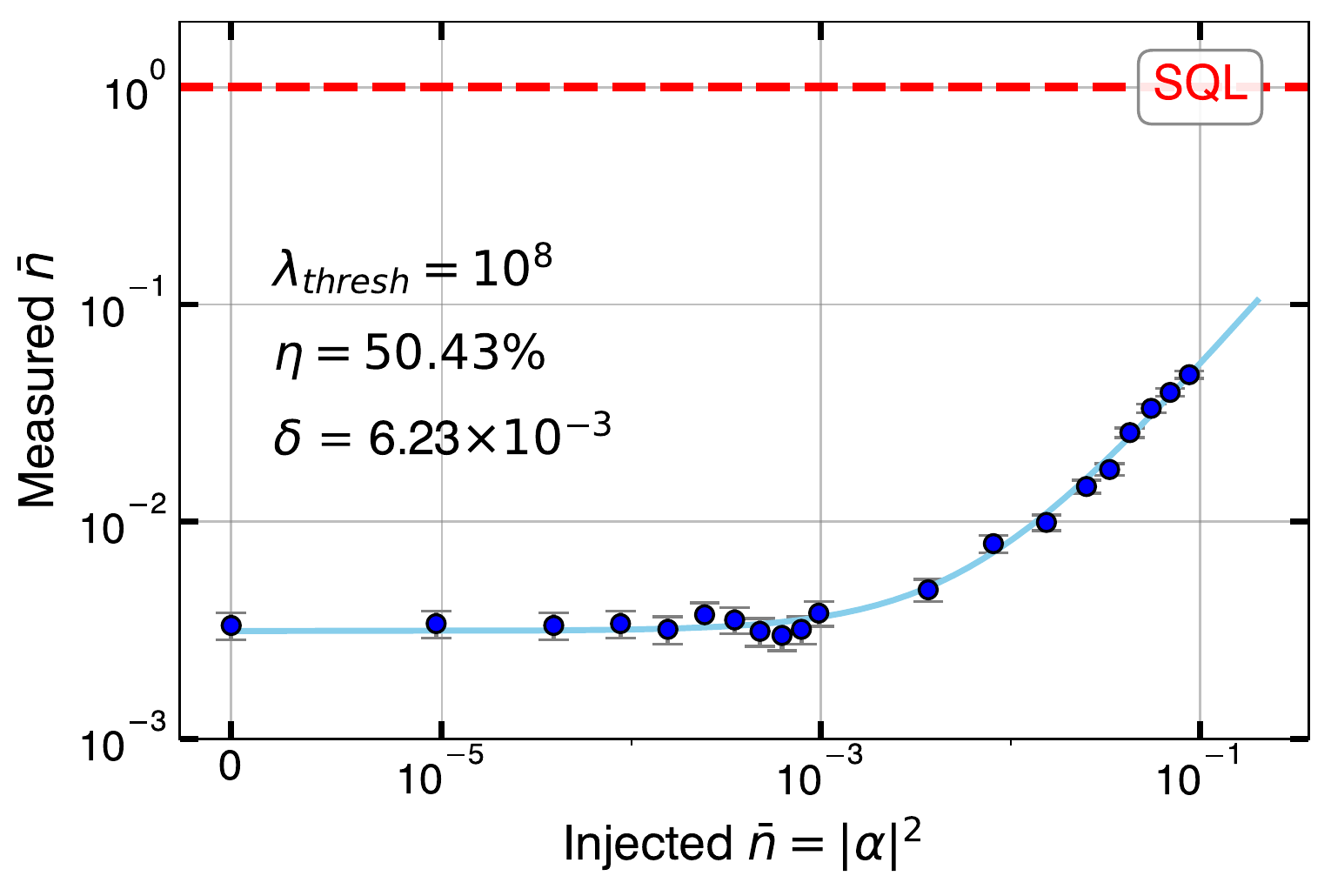}
    }
    \caption{\textbf{Photon counting to measure the cavity background.} Repeated parity measurement technique followed from \cite{Dixit2021} to measure the real photon occupation in the cavity $\bar{n}_{b}^{c} = 6.23 \cdot 10^{-3}$ with detector based errors less than $10^{-8}$.}
    \label{fig:cavity_bkgd_pc}
\end{figure}

The detector efficiency and false positive probability is determined at varying thresholds for detection $\lambda_{\mathrm{thresh}}$. As the detection threshold is increased, more repeated number resolved qubit measurements are required to determine the presence of a photon. This suppresses false positives due to qubit errors but also leads to a decrease in the detector efficiency as events with low likelihood ratio are now rejected. Also, the maximum achievable likelihood ratio decreases with the Fock state prepared in the cavity. Hence, we decided to keep the false positive probability same for comparing different Fock states at the expense of detection efficiency. We performed a single photon counting experiment \cite{Dixit2021} to determine the occupation number of the background photons in the cavity (Fig. \ref{fig:cavity_bkgd_pc}) and chose the threshold such that  $\frac{1}{\lambda_{\mathrm{thresh}}+1} < \bar{n}_{b}^{c}\Rightarrow \lambda_{\rm thresh}=10^{3}$. This measured background translates into a metrological gain of $\mathrm{11.0\, dB}$ below the SQL.

\subsection{Cavity backgrounds}
The number of events which cross $\lambda_{\mathrm{thresh}}$ for cavity prepared in different Fock states are not same, and are subtracted from the $\alpha \neq0$ events to demonstrate the enhancement effect. The number of such events are listed in the table below and are similar to the false positive probability from the fits in Fig. \ref{fig:det_char_stimem}. The number of counts are corrected for the detection efficiency to reflect the actual counts. There is no reason to expect the number of counts would increase with $n$ and we include them in the systematic uncertainties while conducting a dark photon search. In an independent dataset, we observe a similar number of counts as reported in Fig. \ref{fig:cavity_bkgd_dwell}.

\begin{table}[ht]
\begin{center}
\begin{tabular}{ c | c | c | c}
\hline
 $\ket{n}$ &$N_{\rm trials}$ & $N_{\rm bkgd}$ & $\%$ \\
 \hline
 $\ket{0}$ & 114915 & 23 & 0.020\\
 $\ket{1}$ & 111601 & 315 & 0.282\\
 $\ket{2}$ & 113364 & 60 & 0.053\\
 $\ket{4}$ & 108626 & 827 & 0.761\\
\hline
\end{tabular}
\caption{\textbf{Background counts}. Number of background counts $N_{\rm bkgd}$ reported for cavity initialized in different Fock states out of $N_{\rm trials}$.}
\label{table:bkgd_events}
\end{center}
\end{table}

\section{Converting background counts to dark photon exclusion}
As described in the main text, for a coherent source we expect the signal rate to be proportional to the term $a_{0}$. We will derive the expression and compute the kinetic mixing angle which corresponds to $a_{0}$. See Supp. Section in Ref. \cite{Dixit2021} for discussion about dark matter induced signal. 

\subsection*{Kinetic mixing angle exclusion}
For a dark matter candidate on resonance with the cavity frequency ($m_{\mathrm{DM}} c^2 = \hbar \omega_c$), the rate of photons deposited in the cavity prepared in a Fock state $\ket{n}$ by the coherent build up of electric field in time $\tau$ ($\tau < T_{1}^{s}, Q_{\rm DM}/{\nu}$) is given by:

\begin{equation}
    \frac{d N_{\mathrm{HP}}}{dt}  = \frac{U/\omega_s}{\tau} =\frac{1}{2} \frac{E^2 V }{\omega_s} \frac{1}{\tau} = \frac{1}{2} J^2_{\mathrm{DM}} (n+1) \tau^{2} \frac{G V}{\omega_s} \frac{1}{\tau}
    \label{eqn:N_DM}
\end{equation}
The stimulated emission factor appears via the enhancement of magnitude of the electric field generated inside the cavity. The volume of the cavity is $34.5 \times 0.5 \times \mathrm{2.5\, cm^{3}} = \mathrm{43.13\, cm^{3}}$. $\mathcal{G}$ encompasses the total geometric factor of the particular cavity used in the experiment. This includes a factor of $1/3$ due to the dark matter field polarization being randomly oriented every coherence time. For the lowest order mode of the rectangular cavity coupled to the qubit with $\textbf{E} = \mathrm(\frac{\pi x}{l}) \mathrm(\frac{\pi y}{w})\textbf{z}$ the geometric form factor is given by:

\begin{equation}
    G = \frac{1}{3} \frac{\left| \int dV E_z \right|^2}{V \int dV \left| E_z \right|^2} = \frac{1}{3} \frac{2^6}{\pi^4}
\end{equation}

The dark photon generated current is set by the density of dark matter in the galaxy $\rho_{\mathrm{DM}} = \mathrm{0.4 \,GeV / cm^3} = 2\pi \times \mathrm{9.67\times 10^{19} \, GHz/ cm^{3}} $:

\begin{equation}
J^2_{\mathrm{DM}} = 2\epsilon^2 m^4 A'^2 = 2\epsilon^2 m^2 \rho_{\mathrm{DM}}
\label{eqn:J_DM}
\end{equation}
where $\epsilon$ is the kinetic mixing angle between the dark photon and the visible matter. Substituting Eqn. \ref{eqn:J_DM} into Eqn. \ref{eqn:N_DM} yields the signal rate of photons deposited in the cavity by a dark photon dark matter candidate:

\begin{equation}
\frac{d N_{\mathrm{HP}}}{dt} = (n+1)\epsilon^2 \rho_{\mathrm{DM}} m_{\mathrm{DM}} G V \tau
\end{equation}

The total number of photons we expect to be deposited is determined by the photon rate and the integration time ($ N_{\mathrm{\rm trials}}\, \tau$) for each Fock state:

\begin{multline}
    N_{\mathrm{HP}} = \frac{d N_{\mathrm{HP}}}{dt} \times \tau \times N_{\mathrm{\rm trials}} \\ = (n+1)\epsilon^2 \rho_{\mathrm{DM}} m_{\mathrm{DM}} G V \tau^{2} N_{\mathrm{\rm trials}}
    \label{eqn:N_HP}
\end{multline}

Comparing the first term in Eq. \ref{eq:hp_fit_eqn}, we can rewrite the terms to obtain $ \epsilon^{2} = \frac{a_{0}}{\rho_{\mathrm{DM}} m_{\mathrm{DM}} G V}$.

\subsection{Calculating 90\% confidence limit}
\begin{table}[ht]
\begin{center}
\begin{tabular}{ c c c }
\hline
 Expt. Parameter & $\Theta$ & $\sigma_{\Theta}$ \\
 \hline
 $a_{m} $  & $1.9 \times 10^{3} (\mathrm{s^{-2}})$ & $\sigma_{a} = 9.807\times 10^{5} (\mathrm{s^{-2}})$\\
 $\omega_s$ & $\mathrm{5.965\, GHz}$ & $\sigma_{\omega_s}=\mathrm{25\, Hz}$ \\
 $Q_{s}$ & $5.11 \times 10^7$ & $\sigma_{Q_s} = 1.4 \times 10^5$ \\
 $V$ & $\mathrm{43.13\, cm^3}$ & $\sigma_V=\mathrm{1.2\, cm^3}$ \\
 $G$ & $0.002$ & $\sigma_G=0.0002$ \\
\hline
\end{tabular}
\caption[Stimulated emission experimental parameters.]{\textbf{Stimulated emission experimental parameters.} Systematic uncertainties of physical parameters in the experiment must be incorporated in determining the excluded dark photon mixing angle $\epsilon$. The uncertainty in the dark photon (HP) conversion is determined in the previous section. The storage cavity frequency uncertainty is obtained by Ramsey interferometry. The quality factor of the cavity is given by $Q_s = \omega_s T_1^s$ so the uncertainty is calculated as $\sigma_{Q_s}^2 = (\omega_s \sigma_{T_1^s})^2 + (T_1^s \sigma_{\omega_s})^2$. The volume uncertainty is estimated by assuming machining tolerances of 0.005 inches in each dimension. The form a factor uncertainty is estimated from assuming $1\%$ error in the simulated structure. Of the experimental quantities, the DP conversion has the largest systematic uncertainty.}
\label{table:expt_params}
\end{center}
\end{table}

By estimating the strength of the coherent drive in the absence of an external drive and the measured background counts for different Fock states, we perform a dark photon search.  We determine the dark photon mixing angle $\epsilon$ that can be excluded at the 90\% confidence level by using standard error propagation formula. We determine the standard deviation on $\epsilon$ given, we have the error estimates for all the parameters tabulated above. The estimated value of $\epsilon_{0} = 1.6 \times 10^{-15} \pm 3.30 \times 10^{-13}$, dominated by the error on $a_{0}$. We can now set the $90\%$ confidence limit on the kinetic mixing angle term as $\epsilon^{90\%} = \epsilon_{0} + 1.28 \sigma_{\epsilon} = 4.24\times 10^{-13}$. This leads us to exclude, with $90\%$ confidence, dark photon with mixing angle $\epsilon^{90\%}$ greater than $4.24\times10^{-13}$ as shown in Fig. \ref{fig:hp_lim_stim_em}.

\begin{figure}[h]
    \centerline{
    \includegraphics[width=\columnwidth]{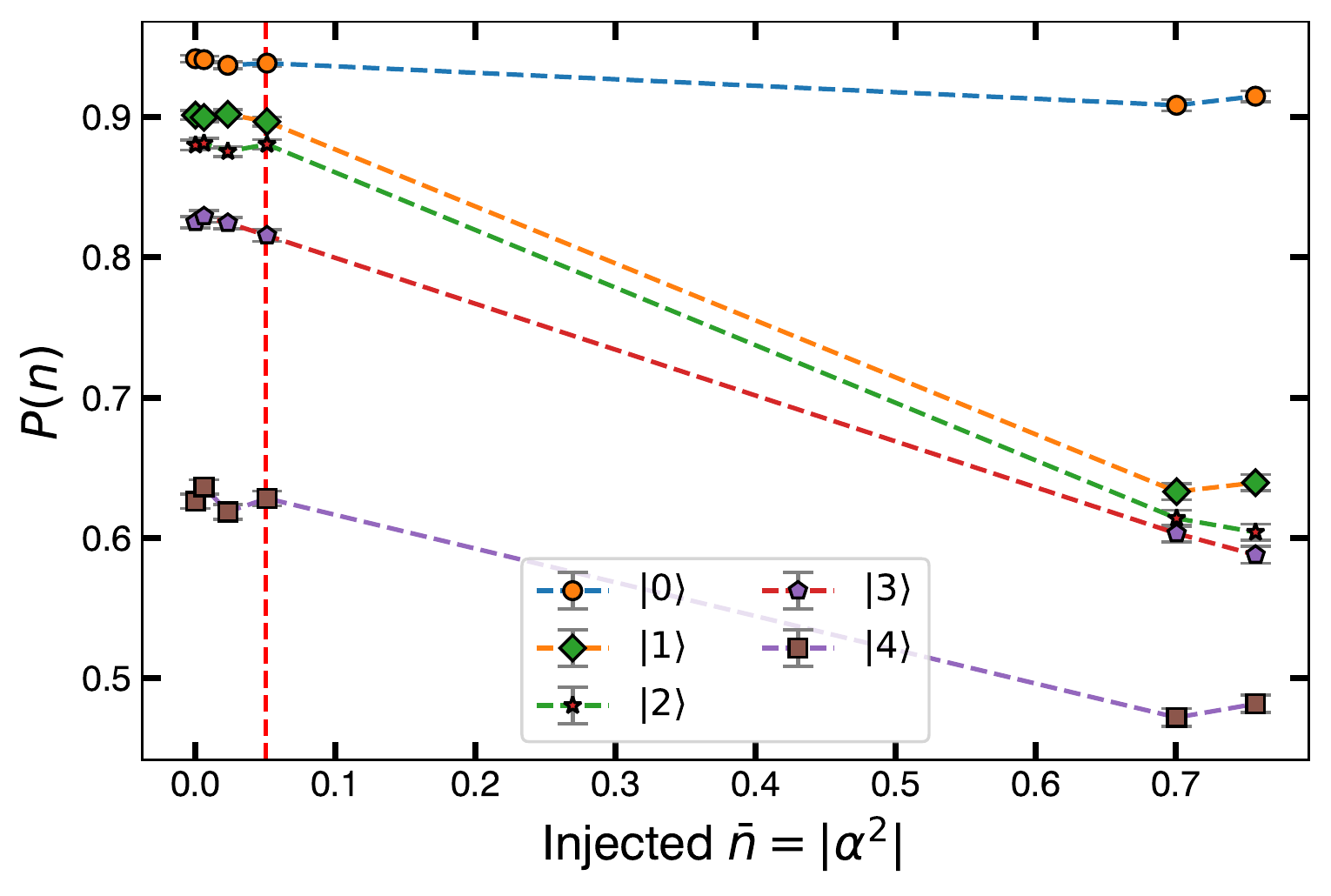}
    }
    \caption{\textbf{Fock state preparation with non-zero occupation in the cavity.} Measuring the Fock state preparation probability as a function of initializing the cavity with varying occupation before the OCT pulses. Red dashed line corresponds to the maximum tolerable injected number of photons where the fidelity changes significantly.}
    \label{fig:max_alpha}
\end{figure}

\section{Dark photon parameter space exclusion}
A dark photon candidate that could result in more detector counts than background counts is constrained by the cavity occupation number which degrades the fidelity of Fock state preparation in the cavity by a significant amount. In order to estimate the same, the cavity is prepared with varied number of mean photons before applying the the OCT pulse. The resultant state is measured with the same procedure as the stimulated emission protocol to compute the fidelity. We observe that the fidelity changes significantly when the mean injected photon number goes above $\bar{n} > 0.05$ shown by the red dashed line in Fig. \ref{fig:max_alpha}. The maximum number of photons sourced from the dark photon which is tolerable before the state preparation is out of control. In principle, we can exclude any value of $\epsilon$ which is above the red curve as it will break the first step of the stimulated emission experiment.
 
\begin{figure}[h]
    \centerline{
    \includegraphics[width=\columnwidth]{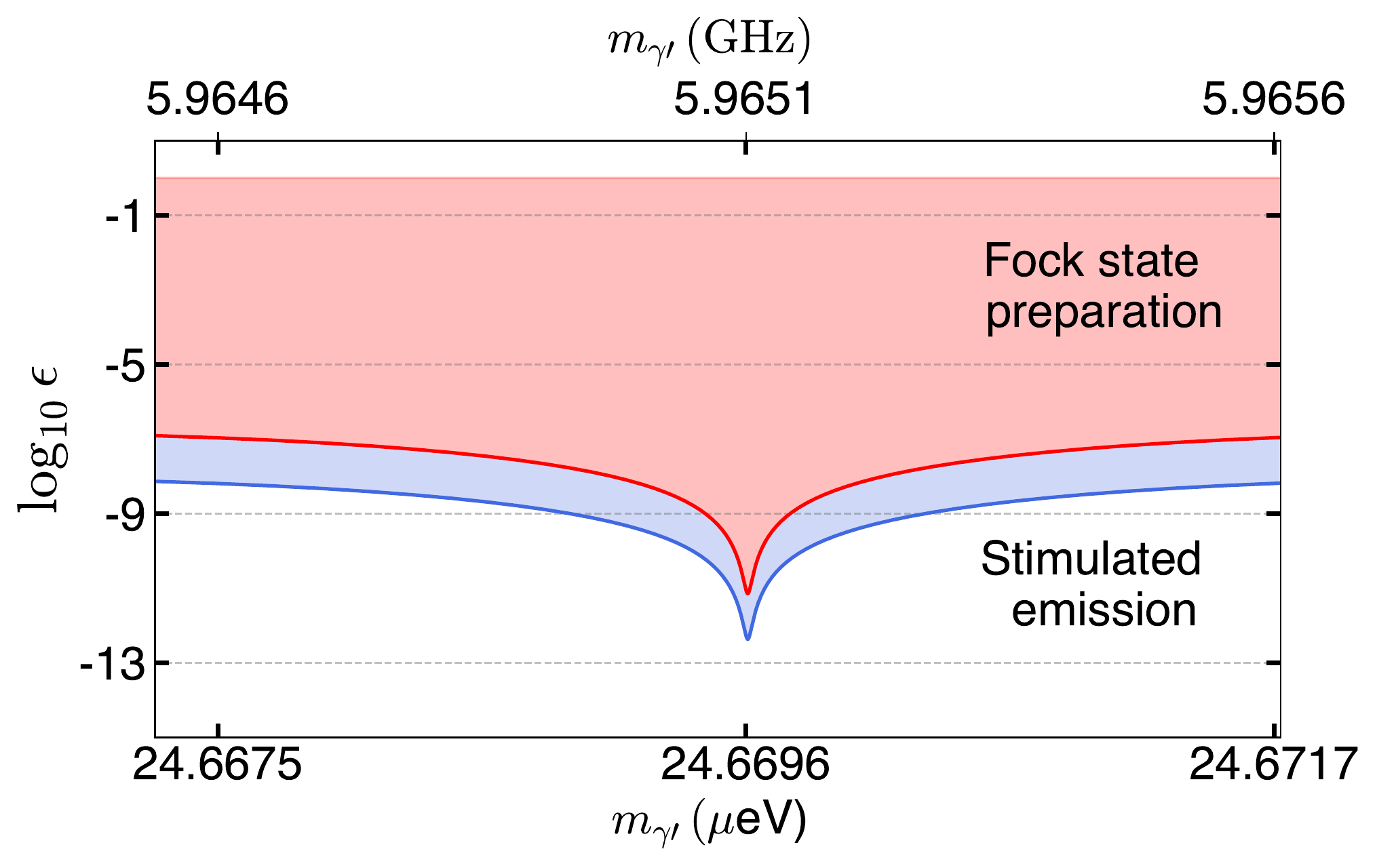}
    }
    \caption[Excluded $\epsilon$ with $\mathrm{m}_{\gamma\prime}$.]{\textbf{Excluded $\epsilon$ with $\mathrm{m}_{\gamma\prime}$.} Shaded regions in the dark photon parameter space of coupling ($\epsilon$) and mass ($m_{\gamma}$) are excluded with $90\%$ confidence. The horizontal extent is set by the bandwidth of the number resolved qubit $\pi$-pulse which is insensitive to any drive outside the band. The vertical limit is set by the minimum $\epsilon$ which would result in dark photon rate greater than the value which would degrade the fidelity of Fock state preparation significantly.}
     \label{fig:hp_lim_stim_em}
\end{figure}

The above calculations assume an infinitely narrow dark matter line. To obtain the excluded region of the dark photon kinetic mixing angle, we must account for the lineshape of the dark matter \cite{Foster2018}. We convolve the dark matter lineshape, characterized by $Q_{\mathrm{DM}} \sim \mathrm{10^6}$, to obtain the region shown in Fig. \ref{fig:hp_lim_stim_em}.

The storage cavity contain an infinite set of discrete resonances each with a unique coupling to the dark matter. We focus only on the lowest order cavity mode that has a non-zero coupling to the dark matter as well as the qubit. In principle, the interactions between any modes and the dark matter could result in additional sensitivity to the dark photon. This would require the mode of interest to have a sufficiently large geometric form factor as well as a resolvable photon number dependent qubit shift. Future dark matter searches could employ structures with multiple resonances to enable multiple simultaneous searches \cite{Chakram2021}.

\bibliographystyle{naturemag}
\bibliography{main.bib}

\begin{thebibliography}{10}
\expandafter\ifx\csname url\endcsname\relax
  \def\url#1{\texttt{#1}}\fi
\expandafter\ifx\csname urlprefix\endcsname\relax\def\urlprefix{URL }\fi
\providecommand{\bibinfo}[2]{#2}
\providecommand{\eprint}[2][]{\url{#2}}

\bibitem{Dixit2021}
\bibinfo{author}{Dixit, A.~V.} \emph{et~al.}
\newblock \bibinfo{title}{Searching for dark matter with a superconducting
  qubit}.
\newblock \emph{\bibinfo{journal}{Phys. Rev. Lett.}}
  \textbf{\bibinfo{volume}{126}}, \bibinfo{pages}{141302}
  (\bibinfo{year}{2021}).
\newblock
  \urlprefix\url{https://link.aps.org/doi/10.1103/PhysRevLett.126.141302}.

\bibitem{Backes2021}
\bibinfo{author}{Backes, K.~M.} \emph{et~al.}
\newblock \bibinfo{title}{A quantum enhanced search for dark matter axions}.
\newblock \emph{\bibinfo{journal}{Nature}} \textbf{\bibinfo{volume}{590}},
  \bibinfo{pages}{238--242} (\bibinfo{year}{2021}).
\newblock \urlprefix\url{https://doi.org/10.1038/s41586-021-03226-7}.

\bibitem{Brubaker2017}
\bibinfo{author}{Brubaker, B.~M.} \emph{et~al.}
\newblock \bibinfo{title}{First results from a microwave cavity axion search at
  $24\text{ }\text{ }\ensuremath{\mu}\mathrm{eV}$}.
\newblock \emph{\bibinfo{journal}{Phys. Rev. Lett.}}
  \textbf{\bibinfo{volume}{118}}, \bibinfo{pages}{061302}
  (\bibinfo{year}{2017}).
\newblock
  \urlprefix\url{https://link.aps.org/doi/10.1103/PhysRevLett.118.061302}.

\bibitem{Tse2019}
\bibinfo{author}{Tse, M.} \emph{et~al.}
\newblock \bibinfo{title}{Quantum-enhanced advanced {LIGO} detectors in the era
  of gravitational-wave astronomy}.
\newblock \emph{\bibinfo{journal}{Physical Review Letters}}
  \textbf{\bibinfo{volume}{123}} (\bibinfo{year}{2019}).
\newblock \urlprefix\url{https://doi.org/10.1103/physrevlett.123.231107}.

\bibitem{Braine2020}
\bibinfo{author}{Braine, T.} \emph{et~al.}
\newblock \bibinfo{title}{Extended search for the invisible axion with the
  axion dark matter experiment}.
\newblock \emph{\bibinfo{journal}{Physical Review Letters}}
  \textbf{\bibinfo{volume}{124}} (\bibinfo{year}{2020}).
\newblock \urlprefix\url{https://doi.org/10.1103/physrevlett.124.101303}.

\bibitem{Kim2023}
\bibinfo{author}{Kim, J.} \emph{et~al.}
\newblock \bibinfo{title}{Near-quantum-noise axion dark matter search at capp
  around $9.5\text{ }\text{ }\mathrm{\ensuremath{\mu}}\mathrm{eV}$}.
\newblock \emph{\bibinfo{journal}{Phys. Rev. Lett.}}
  \textbf{\bibinfo{volume}{130}}, \bibinfo{pages}{091602}
  (\bibinfo{year}{2023}).
\newblock
  \urlprefix\url{https://link.aps.org/doi/10.1103/PhysRevLett.130.091602}.

\bibitem{Chou2022}
\bibinfo{author}{Chou, A.~S.} \emph{et~al.}
\newblock \bibinfo{title}{Snowmass cosmic frontier report}.
\newblock \emph{\bibinfo{journal}{arXiv preprint arXiv:2211.09978}}
  (\bibinfo{year}{2022}).
\newblock \urlprefix\url{https://doi.org/10.48550/arXiv.2211.09978}.
\newblock \eprint{2211.09978v1}.

\bibitem{Tanabashi2018}
\bibinfo{author}{Tanabashi, M.} \emph{et~al.}
\newblock \bibinfo{title}{Review of particle physics}.
\newblock \emph{\bibinfo{journal}{Physical Review D}}
  \textbf{\bibinfo{volume}{98}} (\bibinfo{year}{2018}).
\newblock \urlprefix\url{https://doi.org/10.1103/physrevd.98.030001}.

\bibitem{Rubin1980}
\bibinfo{author}{Rubin, V.~C.}, \bibinfo{author}{Thonnard, N.} \&
  \bibinfo{author}{W.~K., J.~F.}
\newblock \bibinfo{title}{Rotational properties of 21 {SC} galaxies with a
  large range of luminosities and radii, from {NGC} 4605 /r = 4kpc/ to {UGC}
  2885 /r = 122 kpc/}.
\newblock \emph{\bibinfo{journal}{The Astrophysical Journal}}
  \textbf{\bibinfo{volume}{238}}, \bibinfo{pages}{471} (\bibinfo{year}{1980}).
\newblock \urlprefix\url{https://doi.org/10.1086/158003}.

\bibitem{Preskill1983}
\bibinfo{author}{Preskill, J.}, \bibinfo{author}{Wise, M.~B.} \&
  \bibinfo{author}{Wilczek, F.}
\newblock \bibinfo{title}{Cosmology of the invisible axion}.
\newblock \emph{\bibinfo{journal}{Physics Letters B}}
  \textbf{\bibinfo{volume}{120}}, \bibinfo{pages}{127--132}
  (\bibinfo{year}{1983}).
\newblock \urlprefix\url{https://doi.org/10.1016/0370-2693%2883%2990637-8}.

\bibitem{Abbott1983}
\bibinfo{author}{Abbott, L.} \& \bibinfo{author}{Sikivie, P.}
\newblock \bibinfo{title}{A cosmological bound on the invisible axion}.
\newblock \emph{\bibinfo{journal}{Physics Letters B}}
  \textbf{\bibinfo{volume}{120}}, \bibinfo{pages}{133--136}
  (\bibinfo{year}{1983}).
\newblock \urlprefix\url{https://doi.org/10.1016/0370-2693%2883%2990638-x}.

\bibitem{Dine1983}
\bibinfo{author}{Dine, M.} \& \bibinfo{author}{Fischler, W.}
\newblock \bibinfo{title}{The not-so-harmless axion}.
\newblock \emph{\bibinfo{journal}{Physics Letters B}}
  \textbf{\bibinfo{volume}{120}}, \bibinfo{pages}{137--141}
  (\bibinfo{year}{1983}).
\newblock \urlprefix\url{https://doi.org/10.1016/0370-2693%2883%2990639-1}.

\bibitem{Arias2012}
\bibinfo{author}{Arias, P.} \emph{et~al.}
\newblock \bibinfo{title}{{WISPy} cold dark matter}.
\newblock \emph{\bibinfo{journal}{Journal of Cosmology and Astroparticle
  Physics}} \textbf{\bibinfo{volume}{2012}}, \bibinfo{pages}{013--013}
  (\bibinfo{year}{2012}).
\newblock \urlprefix\url{https://doi.org/10.1088/1475-7516/2012/06/013}.

\bibitem{Graham2016}
\bibinfo{author}{Graham, P.~W.}, \bibinfo{author}{Mardon, J.} \&
  \bibinfo{author}{Rajendran, S.}
\newblock \bibinfo{title}{Vector dark matter from inflationary fluctuations}.
\newblock \emph{\bibinfo{journal}{Physical Review D}}
  \textbf{\bibinfo{volume}{93}} (\bibinfo{year}{2016}).
\newblock \urlprefix\url{https://doi.org/10.1103/physrevd.93.103520}.

\bibitem{Sikivie1983}
\bibinfo{author}{Sikivie, P.}
\newblock \bibinfo{title}{Experimental tests of the "invisible" axion}.
\newblock \emph{\bibinfo{journal}{Physical Review Letters}}
  \textbf{\bibinfo{volume}{51}}, \bibinfo{pages}{1415--1417}
  (\bibinfo{year}{1983}).
\newblock \urlprefix\url{https://doi.org/10.1103/physrevlett.51.1415}.

\bibitem{Caves1982}
\bibinfo{author}{Caves, C.~M.}
\newblock \bibinfo{title}{Quantum limits on noise in linear amplifiers}.
\newblock \emph{\bibinfo{journal}{Physical Review D}}
  \textbf{\bibinfo{volume}{26}}, \bibinfo{pages}{1817--1839}
  (\bibinfo{year}{1982}).
\newblock \urlprefix\url{https://doi.org/10.1103/physrevd.26.1817}.

\bibitem{Yamamoto2008}
\bibinfo{author}{Yamamoto, T.} \emph{et~al.}
\newblock \bibinfo{title}{Flux-driven josephson parametric amplifier}.
\newblock \emph{\bibinfo{journal}{Applied Physics Letters}}
  \textbf{\bibinfo{volume}{93}}, \bibinfo{pages}{042510}
  (\bibinfo{year}{2008}).
\newblock \urlprefix\url{https://doi.org/10.1063/1.2964182}.

\bibitem{Eichler2014}
\bibinfo{author}{Eichler, C.} \& \bibinfo{author}{Wallraff, A.}
\newblock \bibinfo{title}{Controlling the dynamic range of a josephson
  parametric amplifier}.
\newblock \emph{\bibinfo{journal}{EPJ Quantum Technology}}
  \textbf{\bibinfo{volume}{1}}, \bibinfo{pages}{2} (\bibinfo{year}{2014}).
\newblock \urlprefix\url{https://doi.org/10.1140/epjqt2}.

\bibitem{Tanay2015}
\bibinfo{author}{Roy, T.} \emph{et~al.}
\newblock \bibinfo{title}{Broadband parametric amplification with impedance
  engineering: Beyond the gain-bandwidth product}.
\newblock \emph{\bibinfo{journal}{Applied Physics Letters}}
  \textbf{\bibinfo{volume}{107}}, \bibinfo{pages}{262601}
  (\bibinfo{year}{2015}).
\newblock \urlprefix\url{https://doi.org/10.1063/1.4939148}.

\bibitem{Martina2019}
\bibinfo{author}{{Esposito, Martina}} \emph{et~al.}
\newblock \bibinfo{title}{Development and characterization of a flux-pumped
  lumped element josephson parametric amplifier}.
\newblock \emph{\bibinfo{journal}{EPJ Web Conf.}}
  \textbf{\bibinfo{volume}{198}}, \bibinfo{pages}{00008}
  (\bibinfo{year}{2019}).
\newblock \urlprefix\url{https://doi.org/10.1051/epjconf/201919800008}.

\bibitem{Caves1981}
\bibinfo{author}{Caves, C.~M.}
\newblock \bibinfo{title}{Quantum-mechanical noise in an interferometer}.
\newblock \emph{\bibinfo{journal}{Physical Review D}}
  \textbf{\bibinfo{volume}{23}}, \bibinfo{pages}{1693--1708}
  (\bibinfo{year}{1981}).
\newblock \urlprefix\url{https://doi.org/10.1103/physrevd.23.1693}.

\bibitem{Lawrie2019}
\bibinfo{author}{Lawrie, B.~J.}, \bibinfo{author}{Lett, P.~D.},
  \bibinfo{author}{Marino, A.~M.} \& \bibinfo{author}{Pooser, R.~C.}
\newblock \bibinfo{title}{Quantum sensing with squeezed light}.
\newblock \emph{\bibinfo{journal}{{ACS} Photonics}}
  \textbf{\bibinfo{volume}{6}}, \bibinfo{pages}{1307--1318}
  (\bibinfo{year}{2019}).
\newblock \urlprefix\url{https://doi.org/10.1021/acsphotonics.9b00250}.

\bibitem{Eickbusch2022}
\bibinfo{author}{Eickbusch, A.} \emph{et~al.}
\newblock \bibinfo{title}{Fast universal control of an oscillator with weak
  dispersive coupling to a qubit}.
\newblock \emph{\bibinfo{journal}{Nature Physics}}
  \textbf{\bibinfo{volume}{18}}, \bibinfo{pages}{1464--1469}
  (\bibinfo{year}{2022}).
\newblock \urlprefix\url{https://doi.org/10.1038/s41567-022-01776-9}.

\bibitem{Hofheinz2008}
\bibinfo{author}{Hofheinz, M.} \emph{et~al.}
\newblock \bibinfo{title}{Generation of fock states in a superconducting
  quantum circuit}.
\newblock \emph{\bibinfo{journal}{Nature}} \textbf{\bibinfo{volume}{454}},
  \bibinfo{pages}{310--314} (\bibinfo{year}{2008}).
\newblock \urlprefix\url{https://doi.org/10.1038/nature07136}.

\bibitem{Wolf_2019}
\bibinfo{author}{Wolf, F.} \emph{et~al.}
\newblock \bibinfo{title}{Motional fock states for quantum-enhanced amplitude
  and phase measurements with trapped ions}.
\newblock \emph{\bibinfo{journal}{Nature Communications}}
  \textbf{\bibinfo{volume}{10}} (\bibinfo{year}{2019}).

\bibitem{Schrodinger1935}
\bibinfo{author}{Schr{\"o}dinger, E.}
\newblock \bibinfo{title}{Die gegenw{\"a}rtige situation in der
  quantenmechanik}.
\newblock \emph{\bibinfo{journal}{Naturwissenschaften}}
  \textbf{\bibinfo{volume}{23}}, \bibinfo{pages}{844--849}
  (\bibinfo{year}{1935}).

\bibitem{Heeres2017}
\bibinfo{author}{Heeres, R.~W.} \emph{et~al.}
\newblock \bibinfo{title}{Implementing a universal gate set on a logical qubit
  encoded in an oscillator}.
\newblock \emph{\bibinfo{journal}{Nature Communications}}
  \textbf{\bibinfo{volume}{8}} (\bibinfo{year}{2017}).
\newblock \urlprefix\url{https://doi.org/10.1038/s41467-017-00045-1}.

\bibitem{Ofek2016}
\bibinfo{author}{Ofek, N.} \emph{et~al.}
\newblock \bibinfo{title}{Extending the lifetime of a quantum bit with error
  correction in superconducting circuits}.
\newblock \emph{\bibinfo{journal}{Nature}} \textbf{\bibinfo{volume}{536}},
  \bibinfo{pages}{441--445} (\bibinfo{year}{2016}).
\newblock \urlprefix\url{https://doi.org/10.1038/nature18949}.

\bibitem{Hu2019}
\bibinfo{author}{Hu, L.} \emph{et~al.}
\newblock \bibinfo{title}{Quantum error correction and universal gate set
  operation on a binomial bosonic logical qubit}.
\newblock \emph{\bibinfo{journal}{Nature Physics}}
  \textbf{\bibinfo{volume}{15}}, \bibinfo{pages}{503--508}
  (\bibinfo{year}{2019}).
\newblock \urlprefix\url{https://doi.org/10.1038/s41567-018-0414-3}.

\bibitem{Campagne_Ibarcq2020}
\bibinfo{author}{Campagne-Ibarcq, P.} \emph{et~al.}
\newblock \bibinfo{title}{Quantum error correction of a qubit encoded in grid
  states of an oscillator}.
\newblock \emph{\bibinfo{journal}{Nature}} \textbf{\bibinfo{volume}{584}},
  \bibinfo{pages}{368--372} (\bibinfo{year}{2020}).
\newblock \urlprefix\url{https://doi.org/10.1038/s41586-020-2603-3}.

\bibitem{Chakram2021}
\bibinfo{author}{Chakram, S.} \emph{et~al.}
\newblock \bibinfo{title}{Seamless high-{$Q$} microwave cavities for multimode
  circuit quantum electrodynamics}.
\newblock \emph{\bibinfo{journal}{Physical Review Letters}}
  \textbf{\bibinfo{volume}{127}} (\bibinfo{year}{2021}).
\newblock \urlprefix\url{https://doi.org/10.1103/physrevlett.127.107701}.

\bibitem{Cahill1969}
\bibinfo{author}{Cahill, K.~E.} \& \bibinfo{author}{Glauber, R.~J.}
\newblock \bibinfo{title}{Density operators and quasiprobability
  distributions}.
\newblock \emph{\bibinfo{journal}{Physical Review}}
  \textbf{\bibinfo{volume}{177}}, \bibinfo{pages}{1882--1902}
  (\bibinfo{year}{1969}).
\newblock \urlprefix\url{https://doi.org/10.1103/physrev.177.1882}.

\bibitem{Koch2007}
\bibinfo{author}{Koch, J.} \emph{et~al.}
\newblock \bibinfo{title}{Charge-insensitive qubit design derived from the
  cooper pair box}.
\newblock \emph{\bibinfo{journal}{Physical Review A}}
  \textbf{\bibinfo{volume}{76}} (\bibinfo{year}{2007}).
\newblock \urlprefix\url{https://doi.org/10.1103/physreva.76.042319}.

\bibitem{Ambegaokar1963}
\bibinfo{author}{Ambegaokar, V.} \& \bibinfo{author}{Baratoff, A.}
\newblock \bibinfo{title}{Tunneling between superconductors}.
\newblock \emph{\bibinfo{journal}{Physical Review Letters}}
  \textbf{\bibinfo{volume}{10}}, \bibinfo{pages}{486--489}
  (\bibinfo{year}{1963}).
\newblock \urlprefix\url{https://doi.org/10.1103/physrevlett.10.486}.

\bibitem{Jaynes_1963}
\bibinfo{author}{{Jaynes}, E.~T.} \& \bibinfo{author}{{Cummings}, F.~W.}
\newblock \bibinfo{title}{Comparison of quantum and semiclassical radiation
  theories with application to the beam maser}.
\newblock \emph{\bibinfo{journal}{Proceedings of the IEEE}}
  \textbf{\bibinfo{volume}{51}}, \bibinfo{pages}{89--109}
  (\bibinfo{year}{1963}).
\newblock \urlprefix\url{https://ieeexplore.ieee.org/document/1443594}.

\bibitem{Schuster2007}
\bibinfo{author}{Schuster, D.~I.} \emph{et~al.}
\newblock \bibinfo{title}{Resolving photon number states in a superconducting
  circuit}.
\newblock \emph{\bibinfo{journal}{Nature}} \textbf{\bibinfo{volume}{445}},
  \bibinfo{pages}{515--518} (\bibinfo{year}{2007}).
\newblock \urlprefix\url{https://doi.org/10.1038/nature05461}.

\bibitem{Brune1990}
\bibinfo{author}{Brune, M.}, \bibinfo{author}{Haroche, S.},
  \bibinfo{author}{Lefevre, V.}, \bibinfo{author}{Raimond, J.~M.} \&
  \bibinfo{author}{Zagury, N.}
\newblock \bibinfo{title}{Quantum nondemolition measurement of small photon
  numbers by rydberg-atom phase-sensitive detection}.
\newblock \emph{\bibinfo{journal}{Physical Review Letters}}
  \textbf{\bibinfo{volume}{65}}, \bibinfo{pages}{976--979}
  (\bibinfo{year}{1990}).
\newblock \urlprefix\url{https://doi.org/10.1103/physrevlett.65.976}.

\bibitem{Gleyzes2007}
\bibinfo{author}{Gleyzes, S.} \emph{et~al.}
\newblock \bibinfo{title}{Quantum jumps of light recording the birth and death
  of a photon in a cavity}.
\newblock \emph{\bibinfo{journal}{Nature}} \textbf{\bibinfo{volume}{446}},
  \bibinfo{pages}{297--300} (\bibinfo{year}{2007}).
\newblock \urlprefix\url{https://doi.org/10.1038/nature05589}.

\bibitem{Heeres2015}
\bibinfo{author}{Heeres, R.~W.} \emph{et~al.}
\newblock \bibinfo{title}{Cavity state manipulation using photon-number
  selective phase gates}.
\newblock \emph{\bibinfo{journal}{Phys. Rev. Lett.}}
  \textbf{\bibinfo{volume}{115}}, \bibinfo{pages}{137002}
  (\bibinfo{year}{2015}).
\newblock
  \urlprefix\url{https://link.aps.org/doi/10.1103/PhysRevLett.115.137002}.

\bibitem{Wang2008}
\bibinfo{author}{Wang, H.} \emph{et~al.}
\newblock \bibinfo{title}{Measurement of the decay of fock states in a
  superconducting quantum circuit}.
\newblock \emph{\bibinfo{journal}{Phys. Rev. Lett.}}
  \textbf{\bibinfo{volume}{101}}, \bibinfo{pages}{240401}
  (\bibinfo{year}{2008}).
\newblock
  \urlprefix\url{https://link.aps.org/doi/10.1103/PhysRevLett.101.240401}.

\bibitem{Nelson2017}
\bibinfo{author}{Leung, N.}, \bibinfo{author}{Abdelhafez, M.},
  \bibinfo{author}{Koch, J.} \& \bibinfo{author}{Schuster, D.}
\newblock \bibinfo{title}{Speedup for quantum optimal control from automatic
  differentiation based on graphics processing units}.
\newblock \emph{\bibinfo{journal}{Phys. Rev. A}} \textbf{\bibinfo{volume}{95}},
  \bibinfo{pages}{042318} (\bibinfo{year}{2017}).
\newblock \urlprefix\url{https://link.aps.org/doi/10.1103/PhysRevA.95.042318}.

\bibitem{Chakram2022}
\bibinfo{author}{Chakram, S.} \emph{et~al.}
\newblock \bibinfo{title}{Multimode photon blockade}.
\newblock \emph{\bibinfo{journal}{Nature Physics}}  (\bibinfo{year}{2022}).
\newblock \urlprefix\url{https://doi.org/10.1038/s41567-022-01630-y}.

\bibitem{Khaneja2005}
\bibinfo{author}{Khaneja, N.}, \bibinfo{author}{Reiss, T.},
  \bibinfo{author}{Kehlet, C.}, \bibinfo{author}{Schulte-Herbrüggen, T.} \&
  \bibinfo{author}{Glaser, S.~J.}
\newblock \bibinfo{title}{Optimal control of coupled spin dynamics: design of
  {NMR} pulse sequences by gradient ascent algorithms}.
\newblock \emph{\bibinfo{journal}{Journal of Magnetic Resonance}}
  \textbf{\bibinfo{volume}{172}}, \bibinfo{pages}{296--305}
  (\bibinfo{year}{2005}).
\newblock \urlprefix\url{https://doi.org/10.1016%2Fj.jmr.2004.11.004}.

\bibitem{Itano1990}
\bibinfo{author}{Itano, W.~M.}, \bibinfo{author}{Heinzen, D.~J.},
  \bibinfo{author}{Bollinger, J.~J.} \& \bibinfo{author}{Wineland, D.~J.}
\newblock \bibinfo{title}{Quantum zeno effect}.
\newblock \emph{\bibinfo{journal}{Physical Review A}}
  \textbf{\bibinfo{volume}{41}}, \bibinfo{pages}{2295--2300}
  (\bibinfo{year}{1990}).
\newblock \urlprefix\url{https://doi.org/10.1103/physreva.41.2295}.

\bibitem{de_Oliveira1990}
\bibinfo{author}{de~Oliveira, F. A.~M.}, \bibinfo{author}{Kim, M.~S.},
  \bibinfo{author}{Knight, P.~L.} \& \bibinfo{author}{Buek, V.}
\newblock \bibinfo{title}{Properties of displaced number states}.
\newblock \emph{\bibinfo{journal}{Physical Review A}}
  \textbf{\bibinfo{volume}{41}}, \bibinfo{pages}{2645--2652}
  (\bibinfo{year}{1990}).
\newblock \urlprefix\url{https://doi.org/10.1103%2Fphysreva.41.2645}.

\bibitem{McDermott2020}
\bibinfo{author}{McDermott, S.~D.} \& \bibinfo{author}{Witte, S.~J.}
\newblock \bibinfo{title}{Cosmological evolution of light dark photon dark
  matter}.
\newblock \emph{\bibinfo{journal}{Physical Review D}}
  \textbf{\bibinfo{volume}{101}} (\bibinfo{year}{2020}).
\newblock \urlprefix\url{https://doi.org/10.1103/physrevd.101.063030}.

\bibitem{Foster2018}
\bibinfo{author}{Foster, J.~W.}, \bibinfo{author}{Rodd, N.~L.} \&
  \bibinfo{author}{Safdi, B.~R.}
\newblock \bibinfo{title}{Revealing the dark matter halo with axion direct
  detection}.
\newblock \emph{\bibinfo{journal}{Physical Review D}}
  \textbf{\bibinfo{volume}{97}} (\bibinfo{year}{2018}).
\newblock \urlprefix\url{https://doi.org/10.1103/physrevd.97.123006}.

\bibitem{Milul2023}
\bibinfo{author}{Milul, O.} \emph{et~al.}
\newblock \bibinfo{title}{A superconducting quantum memory with tens of
  milliseconds coherence time}.
\newblock \emph{\bibinfo{journal}{arXiv preprint arXiv:2302.06442}}
  (\bibinfo{year}{2023}).

\bibitem{Johansson2012}
\bibinfo{author}{Johansson, J.}, \bibinfo{author}{Nation, P.} \&
  \bibinfo{author}{Nori, F.}
\newblock \bibinfo{title}{{QuTiP}: An open-source python framework for the
  dynamics of open quantum systems}.
\newblock \emph{\bibinfo{journal}{Computer Physics Communications}}
  \textbf{\bibinfo{volume}{183}}, \bibinfo{pages}{1760--1772}
  (\bibinfo{year}{2012}).
\newblock \urlprefix\url{https://doi.org/10.1016/j.cpc.2012.02.021}.

\bibitem{Johansson2013}
\bibinfo{author}{Johansson, J.}, \bibinfo{author}{Nation, P.} \&
  \bibinfo{author}{Nori, F.}
\newblock \bibinfo{title}{{QuTiP} 2: A python framework for the dynamics of
  open quantum systems}.
\newblock \emph{\bibinfo{journal}{Computer Physics Communications}}
  \textbf{\bibinfo{volume}{184}}, \bibinfo{pages}{1234--1240}
  (\bibinfo{year}{2013}).
\newblock \urlprefix\url{https://doi.org/10.1016/j.cpc.2012.11.019}.

\bibitem{Pippard1947}
\bibinfo{author}{Pippard, A.~B.} \& \bibinfo{author}{Bragg, W.~L.}
\newblock \bibinfo{title}{The surface impedance of superconductors and normal
  metals at high frequencies ii. the anomalous skin effect in normal metals}.
\newblock \emph{\bibinfo{journal}{Proceedings of the Royal Society of London.
  Series A. Mathematical and Physical Sciences}}
  \textbf{\bibinfo{volume}{191}}, \bibinfo{pages}{385--399}
  (\bibinfo{year}{1947}).
\newblock
  \urlprefix\url{https://royalsocietypublishing.org/doi/abs/10.1098/rspa.1947.0122}.

\bibitem{Reuter1948}
\bibinfo{author}{Reuter, G. E.~H.}, \bibinfo{author}{Sondheimer, E.~H.} \&
  \bibinfo{author}{Wilson, A.~H.}
\newblock \bibinfo{title}{The theory of the anomalous skin effect in metals}.
\newblock \emph{\bibinfo{journal}{Proceedings of the Royal Society of London.
  Series A. Mathematical and Physical Sciences}}
  \textbf{\bibinfo{volume}{195}}, \bibinfo{pages}{336--364}
  (\bibinfo{year}{1948}).
\newblock
  \urlprefix\url{https://royalsocietypublishing.org/doi/abs/10.1098/rspa.1948.0123}.

\bibitem{Zheng_2016}
\bibinfo{author}{Zheng, H.}, \bibinfo{author}{Silveri, M.},
  \bibinfo{author}{Brierley, R.~T.}, \bibinfo{author}{Girvin, S.~M.} \&
  \bibinfo{author}{Lehnert, K.~W.}
\newblock \bibinfo{title}{Accelerating dark-matter axion searches with quantum
  measurement technology} (\bibinfo{year}{2016}).
\newblock \urlprefix\url{https://arxiv.org/abs/1607.02529}.

\bibitem{Sun2014}
\bibinfo{author}{Sun, L.} \emph{et~al.}
\newblock \bibinfo{title}{Tracking photon jumps with repeated quantum
  non-demolition parity measurements}.
\newblock \emph{\bibinfo{journal}{Nature}} \textbf{\bibinfo{volume}{511}},
  \bibinfo{pages}{444--448} (\bibinfo{year}{2014}).
\newblock \urlprefix\url{https://doi.org/10.1038%2Fnature13436}.

\bibitem{Hann2018}
\bibinfo{author}{Hann, C.~T.} \emph{et~al.}
\newblock \bibinfo{title}{Robust readout of bosonic qubits in the dispersive
  coupling regime}.
\newblock \emph{\bibinfo{journal}{Physical Review A}}
  \textbf{\bibinfo{volume}{98}} (\bibinfo{year}{2018}).
\newblock \urlprefix\url{https://doi.org/10.1103/physreva.98.022305}.

\bibitem{Jin2015}
\bibinfo{author}{Jin, X.} \emph{et~al.}
\newblock \bibinfo{title}{Thermal and residual excited-state population in a 3d
  transmon qubit}.
\newblock \emph{\bibinfo{journal}{Physical Review Letters}}
  \textbf{\bibinfo{volume}{114}} (\bibinfo{year}{2015}).
\newblock \urlprefix\url{https://doi.org/10.1103/physrevlett.114.240501}.

\bibitem{Lu1989}
\bibinfo{author}{Lu, N.}
\newblock \bibinfo{title}{Effects of dissipation on photon statistics and the
  lifetime of a pure number state}.
\newblock \emph{\bibinfo{journal}{Physical Review A}}
  \textbf{\bibinfo{volume}{40}}, \bibinfo{pages}{1707--1708}
  (\bibinfo{year}{1989}).
\newblock \urlprefix\url{https://doi.org/10.1103%2Fphysreva.40.1707}.

\end{thebibliography}





\end{document}